\documentclass[pdftex,twocolumn]{article}          % twocolumn

\usepackage[a4paper,left=2cm,right=2cm,top=2.2cm,bottom=2.5cm]{geometry}
\RequirePackage[T1]{fontenc}
\RequirePackage{mathptmx}      % use Times fonts if available on your TeX system

\usepackage{graphicx,subfigure,amsmath,amssymb}
\usepackage[english]{babel}
\usepackage{cuted}

\providecommand{\dif}{\mathrm{d}} \def\d{\dif}
\def\beq{\begin{equation}}
\def\eeq{\end{equation}}
\def\bea{\begin{eqnarray}}
\def\eea{\end{eqnarray}}
\def\nn{\nonumber}
\def\der{|}

\providecommand{\dif}{\mathrm{d}} \def\d{\dif}
\def\nn{\nonumber}

 \def\tens{\mu}

\def\cP{\Pi}
\def\mP{P}

\def\af{\zeta}

\def\der{|}
\def\ISSP{ISEP} % innermost stable string loop position
\def\sun{\odot}

\newcommand{\nnd}{\discretionary{--}{--}{--}}
\newcommand{\Schw}{Schwarzschild}
\newcommand{\dS}{de~Sitter}

\begin{document}

\title{Charged string loops in Reissner-Nordstr\"{o}m black hole background}

\author{
Tursinbay Oteev\\ \texttt{oteevtp@gmail.com} 
\and  
Martin Kolo\v{s}\\ \texttt{martin.kolos@fpf.slu.cz} 
\and 
Zden\v{e}k Stuchl{\'i}k\\ \texttt{zdenek.stuchlik@fpf.slu.cz} 
}

\date{ \it
Institute of Physics and Research Centre of Theoretical Physics and Astrophysics,\\ Faculty of Philosophy \& Science, Silesian University in Opava,\\ Bezru{\v c}ovo n{\'a}m.13, CZ-74601 Opava, Czech Republic\label{addr1}
}

\maketitle

\begin{abstract}
We study the motion of current carrying charged string loops in the Reissner-Nordstr\"{o}m black hole background combining the gravitational and electromagnetic field. Introducing new electromagnetic interaction between central charge and charged string loop makes the string loop equations of motion to be non-integrable even in the flat spacetime limit, but it can be governed by an effective potential even in the black hole background. We classify different types of the string loop trajectories using effective potential approach, and we compare the innermost stable string loop positions with loci of the charged particle innermost stable orbits. We examine string loop small oscillations around minima of the string loop effective potential, and we plot radial profiles of the string loop oscillation frequencies for both the radial and vertical modes. We construct charged string loop quasi-periodic oscillations model and we compare it with observed data from microquasars GRO 1655-40, XTE 1550-564, and GRS 1915+105. We also study the acceleration of current carrying string loops along the vertical axis and the string loop ejection from RN black hole neighbourhood, taking also into account the electromagnetic interaction.
\end{abstract}

%%%%%%%%%%%%%%%%%%%%%%%%%%%%%%%%%%%%%%%%%%%%%%%%%%%%%%%%%%%%%%%%%%%%%%%%%%%%%%%%%

\section{Introduction}

%%%%%%%%%%%%%%%%%%%%%%%%%%%%%%%%%%%%%%%%%%%%%%%%%%%%%%%%%%%%%%%%%%%%%%%%%%%%%%%%%

Detailed studies of relativistic current-carrying string loops moving axisymmetrically along the symmetry axis of Kerr or \Schw\nnd\dS{} black holes appeared currently~\cite{Jac-Sot:2009:PHYSR4:, Kol-Stu:2013:PHYSR4:, Kol-Stu:2010:PHYSR4:}. Tension of such string loops prevents their expansion beyond some radius, while their world-sheet current introduces an angular momentum barrier preventing collapse into the black hole. Such a configuration was also studied in~\cite{Stu-Kol:2012:PHYSR4:, Lar:1994:CLAQG:, Fro-Lar:1999:CLAQG:}. There is an important possible astrophysical relevance of the current-carrying string loops~\cite{Jac-Sot:2009:PHYSR4:} as they could in a simplified way represent plasma that exhibits associated string-like behavior via dynamics of the field lines in the plasma~\cite{Sem-Dya-Pun:2004:Sci:, Chri-Hin:1999:PhRvD:} or due to thin isolated flux tubes of plasma that could be described by an one-dimensional string~\cite{Sem-Dya-Pun:2004:Sci:, Spr:1981:AA:,Cre-Stu:2013:PhRvE:}.

In the previously mentioned articles the string loop was electromagnetically neutral and there was no external electromagnetic field. Motion of electromagnetically charged string loops in combined external gravitational and electromagnetic fields has been recently studied~\cite{Tur-etal:2013:PHYSR4:,Tur-etal:2014:PHYSR4:}. Now we would like to extend such research and we examine dynamic properties of electromagnetically charged and current carrying string loop also in combined electromagnetic and gravitational fields of Reissner-Nordstr\"{o}m background representing a point-like electric charge $Q$ source.
% astrophysical relevance
Our work demonstrates the effect of the black hole charge $Q$ on the string loop dynamic in general; the discussion of the black hole charge relevance is given in the Appendix \ref{appendix1}.  We discuss two astrophysically crucial limiting cases of the dynamics of the charged string loops related to phenomena observed in microquasars: small oscillations around equilibrium radii that can be relevant for the observed quasiperiodic high-frequency oscillations, and strong acceleration of the string loops along the symmetry axis of the black hole - string loop system that can be relevant for creation of jets.

The general dynamics of motion for relativistic current and charge carrying string loop with tension $\mu$ and scalar field $\varphi$ was introduced by~\cite{Lar:1994:CLAQG:} for the spherically symmetric Schwarzschild BH spacetimes, for the Kerr spacetimes it is discussed in \cite{Larsen:1991:PLB:, Lar-Axe:1997:CQG:,Jac-Sot:2009:PHYSR4:, Kol-Stu:2013:PHYSR4:}. General Hamiltonian form for all axially symmetric spacetimes also with electromagnetic field is introduced by~\cite{Lar:1994:CLAQG:}. To show properly how the string loops interact electromagnetically, we will compare charged particle motion with the charged string loop in the same Reissner-Nordstr\"{o}m black hole background, using results already obtained in \cite{Pug-Que-Ruf:2011:PHYSR4:, Pug-Que-Ruf:2017:EPJC:, Bic-Stu-Bal:1989:BAC:, Bal-Bic-Stu:1989:BAC:}. We show that there are similarities in the dynamics of the charged string loops and charged test particles, as the dynamics can be described in both cases by the Hamiltonian formalism with a relatively simple effective potential. There is a fundamental difference in the RN backgrounds: while the test particle motion is regular, the string loop motion has in general chaotic character \cite{Lar:1994:CLAQG:, Kol-Stu:2013:PHYSR4:}, where "islands" of regularity occur only for small oscillations near the string loop stable equilibrium points.

Throughout the present paper we use the spacelike signature $(-,+,+,+)$, and the system of geometric units in which $G = 1 = c$. However, for expressions having an astrophysical relevance we use the constants explicitly. Greek indices are taken to run from 0 to 3.

%%%%%%%%%%%%%%%%%%%%%%%%%%%%%%%%%%%%%%%%%%%%%%%%%%%%%%%%%%%%%%%%%%%%%%%%%%%%%%%%%

\section{Dynamics in spherically symmetric spacetimes}

%%%%%%%%%%%%%%%%%%%%%%%%%%%%%%%%%%%%%%%%%%%%%%%%%%%%%%%%%%%%%%%%%%%%%%%%%%%%%%%%%

Gravitational interaction of the string loop with the central electrically charged black hole occurs through the spherically symmetric Reissner-Nordstr\"{o}m (RN) metric given by the line element expressed in geometric units
\beq \label{SfSymMetrika}
 ds^2=-f(r) dt^2 + f^{-1}(r) dr^2 + r^2 (d\theta^2 + \sin^2\theta d\phi^2),
\eeq
where the metric function reads
\beq
 f(r)=1-\frac{2M}{r}+\frac{Q^2}{r^2}.
\eeq

In the metric function $f(r)$, the parameter $M$ stands for the black hole mass, while $Q$ stands for the black hole charge.
For $0\leq~Q~<M$ the metric (\ref{SfSymMetrika}) describes black hole with two event horizons, located at
\beq
r_{\rm h\pm} = M\pm\sqrt{M^2-Q^2}, \label{RNhorizon}
\eeq
for $Q=M$ there is just one degenerate event horizon solution, for $Q>M$ we have naked singularity without horizons. Hereafter in this paper we will use for simplicity the system of units in which the mass of the black hole $M=1$, i.e., we express the related quantities in units of the black hole mass.

In order to clearly show the trajectory of string loops, it is useful to use the Cartesian coordinates $x,y,z$ related to the \Schw{} coordinates
\beq \label{ccord}
 x = r \sin(\theta) \cos(\phi), \quad y = r \sin(\theta) \sin(\phi), \quad z = r \cos(\theta).
\eeq

The electromagnetic field related to the Reissner-Nordstr\"{o}m (RN) metric is given by the covariant electromagnetic four-vector potential $A_\alpha$ \cite{Mis-Tho-Whe:1973:Gra:} that takes the simple form
\beq \label{EMfield}
A_\alpha = \frac{Q}{r} \, \left(-1,0,0,0 \right).
\eeq
Recall that electromagnetic interaction is much more stronger than the gravitational interaction - between electron and proton for example the gravitational interaction is physically irrelevant. Therefore the central charge $Q$ will significantly electrically interact with the charged string loop even when the black hole charge $Q$ is too small to make relevant contribution to the metric (\ref{SfSymMetrika}).

We thus examine different physically relevant situations according to the gravitational/electric field strength ratio:
\begin{description}
\item[Flat] There is no black hole and hence no gravitational interaction. The electric field of the charge $Q$ is so weak, that it will not contribute to the metric. We will use the flat metric (\ref{SfSymMetrika}), with $M=0,Q=0$, while the electromagnetic interaction will be given by (\ref{EMfield}). Discussed in section \ref{loopFLAT}.
\item[RN] There will be black hole, with the gravitational field influenced by the strong electromagnetic field. We will use full RN metric (\ref{SfSymMetrika}) with electromagnetic interaction given by (\ref{EMfield}). Discussed in section \ref{loopRN}.
\end{description}
The relevance of the individual three cases for realistic values of RN black hole metric/string loop, all string loop quantities and their dimensions in physical units, will be discussed in detail in the Appendix \ref{appendix1}.

%%%%%%%%%%%%%%%%%%%%%%%%%%%%%%%%%%%%%%%%%%%%%%%%%%%%%%%%%%%%%%%%%%%%%%%%%%%%%%%%%
\subsection{Hamiltonian formalism for charged particle motion}

The dynamics of axially symmetric charged current carrying string loops can be enlightened by comparison with charged test particle motion, as both these dynamics can be formulated in the framework of Hamiltonian formalism. Recall that evolution of axisymmetric string loops adjusted to axisymmetric backgrounds can be represented by evolution of a single point of the string \cite{Stu-Kol:2012:PHYSR4:, Kol-Stu:2013:PHYSR4:}.

We can also compare  electromagnetic forces acting on charge test particles or string loops. Since the motion of a charged test particle in the RN black hole background has been intensively studied in literature \cite{Pug-Que-Ruf:2011:PHYSR4:, Pug-Que-Ruf:2017:EPJC:, Bic-Stu-Bal:1989:BAC:, Bal-Bic-Stu:1989:BAC:}, we will give just short summary.

Motion of a charged particle with mass $m$ and charge $q$ is given by the Hamiltonian \cite{Mis-Tho-Whe:1973:Gra:}
\beq \label{particleHAM}
H_{\rm p} =  \frac{1}{2} g^{\alpha\beta} (\cP_\alpha - q A_\alpha)(\cP_\beta - q A_\beta) + \frac{1}{2} \, m^2,
\eeq	
where mechanical, $\mP^\mu$, and canonical, $\cP^\mu$, momenta are related as
\beq	
\mP^\mu = m U^\mu = m \frac{\d x^{\mu}}{\d \tau} = \cP^\mu - q A^\mu.
\eeq

Due to the spherical symmetry of the RN background (\ref{SfSymMetrika}), the charged particles move in central planes only. For a single particle the central plane can be chosen as the equatorial plane. Since the Hamiltonian (\ref{particleHAM}) does not contain coordinate $\phi$ (axial symmetry) and coordinate $t$ explicitly, two constants of motion exist - particle energy $E =-\cP_t$, and particle axial angular momentum $L=\cP_\phi$. Now one can write the Hamiltonian of the charged particle equatorial motion in the form
\beq
 H = \frac{1}{2} f(r) \mP_r^2 - \frac{1}{2} \frac{1}{f(r)} \left( E - \frac{q Q}{r} \right)^2  + \frac{1}{2}\frac{L^2}{r^2} + \frac{1}{2} m^2,
\label{HAMp1}
\eeq
Using the $H=0$ condition, we obtain immediately equation of the radial motion in the form
\beq
 m^2 \left(\frac{\d r}{\d \tau}\right)^2 = \left( E - \frac{q Q}{r} \right)^2  - f(r) \left( m^2 + \frac{L^2}{r^2}\right),
\eeq
corresponding to the motion in 1D effective potential determining the turning points of the radial motion where $ \d r / \d \tau = 0 $.

% $\mP_r=0$ condition

% angular velocity $\omega = \mP^\phi/m$

%%%%%%%%%%%%%%%%%%%%%%%%%%%%%%%%%%%%%%%%%%%%%%%%%%%%%%%%%%%%%%%%%%%%%%%%%%%%%%%%%
\subsection{Hamiltonian formalism for relativistic string loop}

Dynamics of relativistic, charged, current carrying string is described by the action $S$ with Lagrangian $\mathcal{L}$ \cite{Lar:1993:CLAQG:,Lar:1994:CLAQG:}
\bea\label{akceEM}
S&=&\int\mathcal{L}\ d\sigma d\tau,\nn\\
\mathcal{L}&=&-\tens \sqrt{-h} - \frac{1}{2} \sqrt{-h} h^{ab} (\varphi_{|a}+A_a)(\varphi_{|b}+A_b),
\eea

where $A_{a}=A_\gamma X^\gamma_{|a}$. The string worldsheet is described by the spacetime coordinates $X^{\alpha}(\sigma^{a})$ with $\alpha = 0,1,2,3$ given as functions of two worldsheet coordinates $\sigma^{a}$ with $a = 0,1$. This implies induced metric on the worldsheet in the form
\beq
h_{ab}= g_{\alpha\beta}X^\alpha_{\der a}X^\beta_{\der b},
\eeq
where $\Box_{\der a} = \partial \Box /\partial a$. The string current localized on the 2D worldsheet is described by a scalar field $\varphi({\sigma^a})$. The 2D worldsheet with coordinates $\tau,\sigma$ is immersed into 4D metrics with coordinates $t,r,\theta,\phi$ using
\beq
 X^\alpha(\tau,\sigma) = (t(\tau),r(\tau),\theta(\tau),\sigma). \label{strcoord}
\eeq
The action (\ref{akceEM}) is inspired by an effective description of superconducting strings representing topological defects occurring in the theory with multiple scalar fields undergoing spontaneous symmetry breaking~\cite{Wit:1985:NuclPhysB:, Vil-She:1994:CSTD:} and can be used as effective description of current created by bosons or fermions on superconducting string. Contrary to the formalism used in ~\cite{Jac-Sot:2009:PHYSR4:, Kol-Stu:2013:PHYSR4:}, we rescale scalar field $\varphi \rightarrow \varphi/2$. First part of (\ref{akceEM}) is classical Nambu---Goto string action for string with tension $\tens$ only, second part describes interaction of scalar field $\varphi$ with four-potential $A_\alpha$ of electromagnetic field.

In the conformal gauge, the equation of motion of the scalar field, given by the variation of the action (\ref{akceEM}) against field $\varphi$, reads
\beq\label{PHIevolution}
 \left[ \sqrt{-h}h^{ab} (\varphi_{|a}+A_a) \right]_{|b} = 0.
\eeq
The assumption of axisymmetry implies $\varphi_{|\sigma\sigma} = 0 $ and $A_{\phi}=A_{\sigma} \neq A_{\sigma}(\phi)$, from (\ref{PHIevolution}) we have conserved quantities $\Omega$ and $n$, given by
\beq \label{Jdef}
 \Omega = \varphi_{|\tau} + A_\tau, \quad n = \varphi_{|\sigma},
\eeq

Varying the action (\ref{akceEM}) with respect to the induced metric $h_{ab}$, we obtain the worldsheet stress-energy tensor density (being of density weight one with respect to worldsheet coordinate transformations)
\bea
\Sigma^{\tau\tau}&=&\frac{\Omega^2+(n+A_{\phi})^2}{g_{\phi\phi}} + \mu,\nn\\
\Sigma^{\sigma\sigma}&=&\frac{\Omega^2+(n+A_{\phi})^2}{g_{\phi\phi}} - \mu,\nn\\
\Sigma^{\sigma\tau}&=&\frac{-\Omega(n+A_{\phi})}{g_{\phi\phi}}.
\eea

The contribution from the string tension $\mu > 0$ gives a positive energy density and a negative pressure (tension). The current contribution is traceless, due to the conformal invariance of the action - it can be considered as a $1+1$ dimensional massless radiation fluid with positive energy density and equal pressure ~\cite{Jac-Sot:2009:PHYSR4:}.

Electromagnetic properties of the charged circular string loop are obtained by varying the action (\ref{akceEM}) with respect to the four-potential $A_\alpha$:
\beq \label{stringJQ}
 J^{\mu} = \frac{\delta \mathcal{L}}{\delta A_{\mu}} = - \rho X^\mu_{|\tau} + j X^\mu_{|\sigma}, \qquad  \frac{\partial \rho}{\partial \tau} = \frac{\partial j}{\partial \sigma},
\eeq
where the string loop electric current is $j=n + A_{\phi}$ and string loop electric charge density is $\rho=\Omega$ ~\cite{Lar:1993:CLAQG:}. % Total charge of the string loop $q$ is given by $q = 2 \pi \Omega$.

Varying the action (\ref{akceEM}) with respect to $X^\mu$ implies equations of motion in the form
\beq
 \frac{\rm D}{\d\tau} \cP^{(\tau)}_\mu + \frac{\rm D}{\d\sigma} \cP^{(\sigma)}_\mu = 0,
\eeq
where the string loop momenta are defined by the relations
\bea
\cP^{(\tau)}_\mu \equiv \frac{\partial \mathcal{L}}{\partial \dot{X}^\mu}&=&\Sigma^{\tau a} g_{\mu \lambda} X^\lambda_{|a} + \Omega A_\mu,\nn\\
\cP^{(\sigma)}_\mu \equiv \frac{\partial \mathcal{L}}{\partial {X'}^\mu}&=&\Sigma^{\sigma a} g_{\mu \lambda} X^\lambda_{|a} - (n+A_{\phi}) A_\mu.
\eea

Defining affine parameter $\af$, related to the worldsheet coordinate $\tau$ by the transformation
\beq
 \d\tau = {\Sigma^{\tau\tau}}  \d\af ,
\eeq
we can define for the string loop dynamics define the Hamiltonian
\beq \label{HamEM}
 H = \frac{1}{2} g^{\alpha\beta} (\cP_\alpha - \Omega A_\alpha)(\cP_\beta - \Omega A_\beta) + \frac{1}{2} g_{\phi\phi} \left[(\Sigma^{\tau\tau})^2 - (\Sigma^{\tau\sigma})^2 \right]
\eeq
and the related Hamilton equations
\beq \label{Ham_eqEM}
\mP^\mu\equiv\frac{\d X^\mu}{\d\af} = \frac{\partial H}{\partial \cP_\mu}, \quad \frac{\d\cP_\mu}{\d\af} = - \frac{\partial H}{\partial X^\mu}.
\eeq
From the first equation in (\ref{Ham_eqEM}) we obtain relation between the canonical $\cP^\mu$ and mechanical momenta $\mP^\mu$ in the form
\beq
 \mP^\mu = \cP^\mu - \Omega A^\mu.
\eeq

%%%%%%%%%%%%%%%%%%%%%%%%%%%%%%%%%%%%%%%%%%%%%%%%%%%%%%%%%%%%%%%%%%%%%%%%%%%%%%%%%
\subsection{Conserved quantities and effective potential for string loop dynamics}

Now we restrict our study to the string loop dynamics in the RN black hole background using metric (\ref{SfSymMetrika}) and electromagnetic four-potential (\ref{EMfield}). The metric (\ref{SfSymMetrika}) does not depend on coordinates $t$ (static) and $\phi$ (axial symmetry) and only one nonzero covariant component of the electromagnetic four-vector potential is $A_t$ (\ref{EMfield}). Such symmetries imply existence of conserved quantities during string motion - string energy $E$ and string axial angular momentum $L$, determined by the relations
\bea\label{Cmotion}
-E&=&\cP_t=\mP_t+\Omega A_t,\nn\\
L&=&\cP_\phi=g_{\phi\phi}\Sigma^{\tau\sigma}+\Omega A_\phi\nn\\
&=&-\Omega n=-2J^2\omega\sqrt{1-\omega ^2}.
\eea

The string loop does not rotate in \Schw{} coordinates, $\d X^\phi / \d\af =0$ - see eq. (\ref{strcoord}), but the string loop has a non-zero angular momentum generated completely by the scalar field living on the string loop. Instead of string loop electric charge $\Omega$ and current $n$ (\ref{stringJQ}), we can introduced new conserved quantities - "angular parameter" $J$ and "charge parameter" $\omega$, given by
\beq \label{JOdef}
 J^2 \equiv \frac{\Omega^2 + n^2}{2}, \qquad \omega \equiv \frac{1}{\sqrt{2}} \frac{\Omega}{J}.
\eeq
The parameter $J\geq~0$ has simple interpretation as combined magnitude of charge and current on the string loop, and as we will see later, $J$ is a generator of the centrifugal force, acting against contraction caused by the string loop tension $\mu$. Note that even though string loop is not rotating mechanically, $J$ parameter is acting as an angular momentum due to internal properties of the string. For this reason we call it angular momentum $J$ parameter. Further, the new parameter $\omega$ is string loop charge $\Omega$ rescaled by parameter $J$, such that $-1\leq\omega\leq~1$. We can distinguish three limiting cases of parameter $\omega$:
\begin{description}
\item[$\omega=-1$] There is no electric current on the string loop, $n=0$, only negative electric charge $\Omega<0$ uniformly distributed along the loop. Since we consider the central object charge $Q>0$ to be positive, there acts an electromagnetic attractive force between the central object and the string loop.
\item[$\omega=0$] There is no charge on the string loop, $\Omega=0$, only current $n$, and there is no electromagnetic interaction between the string loop and the central object electric charge $Q$. The black hole charge $Q$ can affect the string loop dynamic only through changes in (\ref{SfSymMetrika}) metric. This case was already studied in \cite{Stu-Kol:2012:JCAP:} for the so called "tidal charge" black hole scenario.
\item[$\omega=1$] There is no electric current on the string loop, $n=0$, only positive electric charge $\Omega>0$ uniformly distributed along the loop. there will be electromagnetic repulsive force between the central object and the string loop.
\end{description}
The electric force between the central object with charge $Q$ is attractive for $-1\leq\omega<0$, while it is repulsive for $0<\omega\leq~1$. We will focus on string loop dynamics for $\omega\in\{-1,0,1\}$ limiting values, and we will assume the string loop behaviour for another value of $\omega$, will be combination of the limiting values. It is interesting that for all three limiting cases $\omega\in\{-1,0,1\}$, the string loop angular momentum $L$ is zero (\ref{Cmotion}).

The string dynamics depends on the $J$ parameter (\ref{Jdef}) through the worldsheet stress-energy tensor. Using the two constants of motion (\ref{Cmotion}), we can rewrite the Hamiltonian (\ref{HamEM}) into the form related to the $r$ and $\theta$ momentum components
\bea \label{HamHam}
H &=& \frac{1}{2} g^{rr} \mP_r^2 + \frac{1}{2} g^{\theta\theta} \mP_\theta^2 + \frac{1}{2} g^{tt} \left( E + \Omega A_{t} \right)^2 \nn\\
&& +\frac{1}{2} g_{\phi\phi} \left(\frac{J^2}{g_{\phi\phi}} + \mu \right)^2.
\eea

Assuming $\mu>0$, one can also express all quantities in the terms of string loop tension $\mu$, divide the whole Hamiltonian (\ref{HamHam}) with $\mu$, and hence get rid of extra parameter $\mu$ with transformations like $ E \rightarrow E/\sqrt{\mu}$ and $J \rightarrow J/\sqrt{\mu}$. Hence, in all following equations, we will take $\mu=1$ and we will discuss the string loop quantities and their dimensions in physical units in the Appendix \ref{appendix1}.

The equations of motion (\ref{Ham_eqEM}) following from the Hamiltonian (\ref{HamHam}) are very complicated and can be solved only numerically in general case, although there exist analytical solutions for simple cases of the motion in the flat or de~Sitter spacetimes ~\cite{Kol-Stu:2010:PHYSR4:}. However, we can tell a lot about the string loop dynamics even without solving the equation of motion (\ref{Ham_eqEM}) by studying properties of the effective potential governing the turning points of the string loop motion that is implied by the Hamiltonian. It is useful to express the Hamiltonian (\ref{HamHam}) in the form
\beq \label{HpartDP}
 H = H_{\rm D} + H_{\rm P},
\eeq
where we split $H$ into "dynamical" $H_{\rm D}$ and "potential" $H_{\rm P}$ parts. The "dynamical" part $H_{\rm D}$ contains all terms with momenta $\cP_\alpha$ or $\mP_\alpha$, while the "potential" part $H_{\rm P}$ depends on the coordinates and conserved quantities only.

The positions where a string loop has zero velocity ($H_{\rm D}=0$ and $\dot{r}=0, \dot{\theta}=0$) forms a boundary for the string motion. Since the total Hamiltonian is zero, $H=0$, the potential part of the Hamiltonian is also zero $H_{\rm D}=0$ at the boundary points with zero velocity. This will allow us to to express string loop energy $E$ at the turning (boundary) points in the form
\beq \label{StringEnergy}
E = V_{\rm eff} (r,\theta) \equiv \sqrt{ -g_{tt} g_{\phi\phi} } \left( \frac{J^2}{g_{\phi\phi}} + 1 \right) -\Omega A_{t},
\eeq
where we define the effective potential function $V_{\rm eff}(r,\theta)$. The condition (\ref{StringEnergy}), for energy $E$, creates an unbreakable boundary (curve) in the $x$-$z$ plane, restricting the string loop motion. In the previous works \cite{Jac-Sot:2009:PHYSR4:,Kol-Stu:2013:PHYSR4:}, the term "energy boundary function" was used for the effective potential, $E_{\rm{b}}(r,\theta)=V_{\rm eff}(r,\theta)$.

Stationary points of the effective potential function $V_{\rm eff}(x,z)$ are determined by two conditions
\beq \label{extr}
(V_{\rm eff})_{x}' = 0  \qquad (V_{\rm eff})_{z}' = 0,
\eeq
where the prime $()_{m}'$ denotes derivation with respect to the coordinate $m$. In order to determine character of the stationary points at $(x_{\rm e},z_{\rm e})$ given by the stationarity conditions (\ref{extr}), i.e., whether we have a maximum ("hill") or minimum ("valley") of the effective potential function $V_{\rm eff}(x,z)$, we have to examine additional conditions
\bea\label{extr_Veff1}
[(V_{\rm eff})_{zz}'' (V_{\rm eff})_{xx}'' - (V_{\rm eff})_{zx}'' (V_{\rm eff})_{xz}''](x_{\rm e},z_{\rm e}) > 0,
\eea
\bea\label{extr_Veff2}
[(V_{\rm eff})_{zz}''] (x_{\rm e},z_{\rm e}) \,\,\, < 0 \, \mathrm{(max)} \,\, > 0 \, \mathrm{(min)}.
\eea

The curve $E=V_{\rm eff}(x,z)$, forming unbreakable energetic boundary for the string loop motion, can be open in the $x$-direction in the equatorial plane ($z=0$), allowing the string loop to move towards horizon and be captured by the black hole. The energetic boundary can be open in $z$-direction, allowing the string loop to escape to infinity from the black hole neighbourhood.

%%%%%%%%%%%%%%%%%%%%%%%%%%%%%%%%%%%%%%%%%%%%%%%%%%%%%%%%%%%%%%%%%%%%%%%%%%%%%%%%%

\section{String loop in combined electric and gravitational field}

%%%%%%%%%%%%%%%%%%%%%%%%%%%%%%%%%%%%%%%%%%%%%%%%%%%%%%%%%%%%%%%%%%%%%%%%%%%%%%%%%

\begin{figure*}
\includegraphics[width=\hsize]{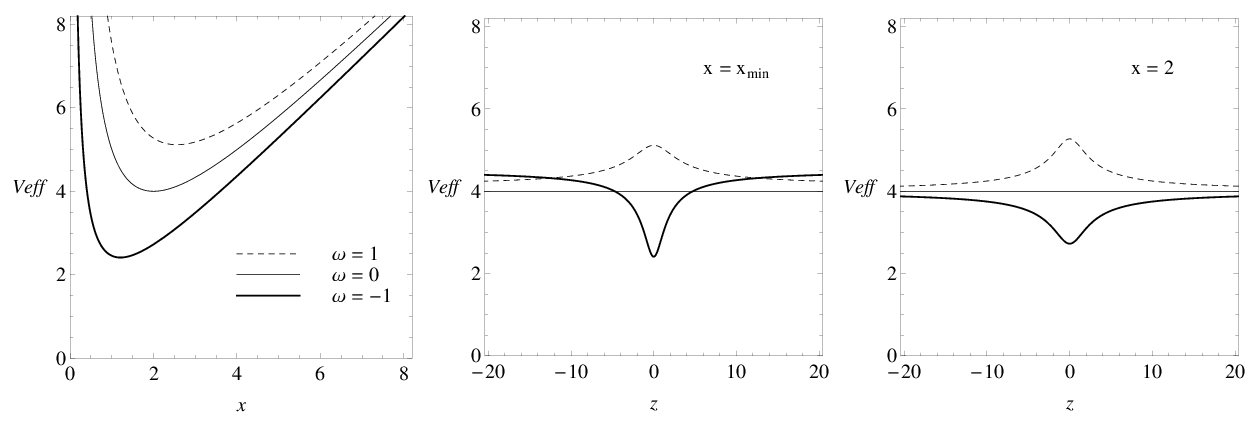}
\caption{\label{RNflat} The $x$ and $z$ sections of the effective potential $V_{\rm eff}(x,z)$ for the flat spacetime. We use parameter $J=2$ and we set $Q=0.9$. Dashed curves correspond to the $\omega=1$ case, solid to the $\omega=0$, thick to the $\omega=-1$. The $x$ section of $V_{\rm eff}$ is taken at $z=0$, while $z$ section is taken at corresponding minima $x=x_{\rm min}$ (middle fig.) or at $x=2$ (right fig.).}
\end{figure*}

%%%%%%%%%%%%%%%%%%%%%%%%%%%%%
\begin{figure*}
\includegraphics[width=\hsize]{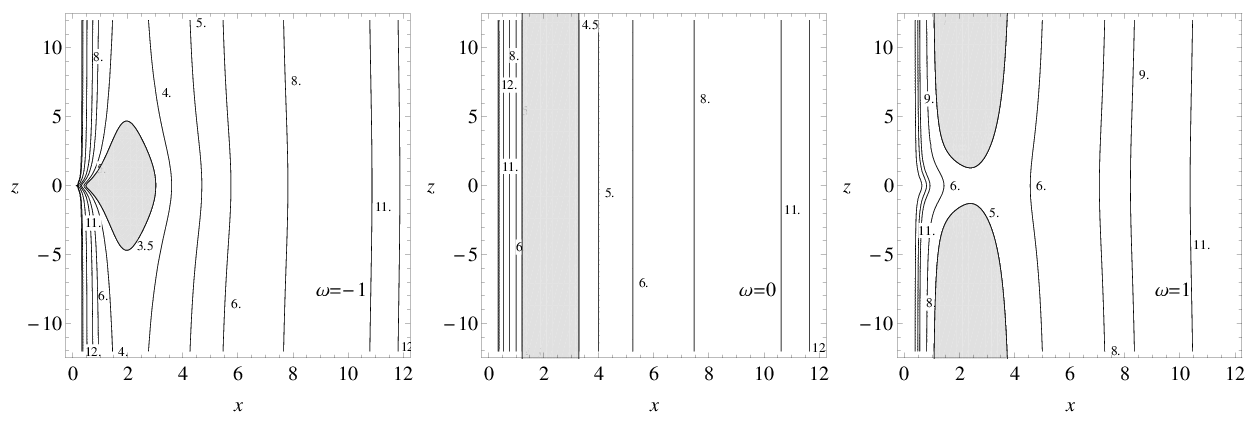}
\caption{\label{RN3D} Energy boundary function $E_{\rm b} (x,z)$ for the flat spacetime. We use parameters $J=2, Q=0.9$. For Fig. $\omega_1 = -1$ we have string energy $E = 3.5$ (left), for $\omega_1 = 0$ $E = 4.5$ (middle) and for $\omega_1 = 1$ $E = 5$ (right)}
\end{figure*}
%%%%%%%%%%%%%%%%%%%%%%%%%%%%%

\begin{figure*}
\includegraphics[width=\hsize]{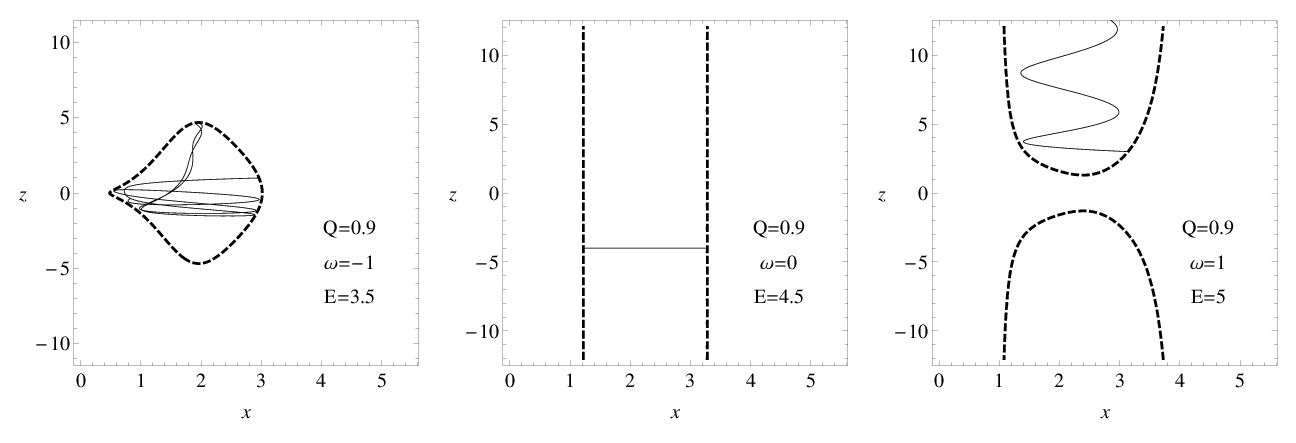}
\caption{\label{FlatTra} Energy boundaries and the trajectories of the string loop with parameters $J=2, Q=0.9$. For Fig. $\omega_1 = -1$ we have string energy $E = 3.5$ (left), for $\omega_1 = 0$ $E = 4.5$ (middle) and for $\omega_1 = 1$ $E = 5$ (right)}
\end{figure*}

%%%%%%%%%%%%%%%%%%%%%%%%%%%%%%%%%%%%%%%%%%%%%%%%%%%%%%%%%%%%%%%%%%%%%%%%%%%%%%%%%
\subsection{Charged string loop in flat spacetime \label{loopFLAT}}

We discuss the flat spacetime case separately, as establishing the flat space limit requires $M=0$ and $Q=0$ simultaneously, but this means vanishing of the electromagnetic field.

We can use cylindrical coordinates $(t,x,z,\phi)$ in flat spacetime, and compare string loop Hamiltonian with Hamiltonian for particle on circular geodesic - this will be very helpful for exploring the situation and for identification of acting forces. In the electrostatic field of point charge $Q$ (\ref{EMfield}) we have for the "potential", $H_{\rm P}$, parts of the Hamiltonian determining the charged particle and charged string loop motion, the simple expressions
\bea\label{HstringEL}
H_{\rm particle}&=&- \frac{1}{2} \left( E - \frac{q Q}{r} \right)^2 + \frac{1}{2} m^2  + \frac{L^2}{x^2},\nn\\
H_{\rm string}&=&- \frac{1}{2} \left( E - \frac{\Omega Q}{r} \right)^2 + \frac{1}{2} \left(\tens x + \frac{J^2}{x} \right)^2.
\eea

While particle can move only in the central plane, taken to be equatorial plane for simplicity, and particle motion remain regular, the string loop can move also outside the equatorial plane, and string loop dynamics is generally chaotic.
In Hamiltonian (\ref{HstringEL}) we clearly see radial Coulombic force $ \sim \Omega Q /r^2$, acting on the element of string loop with electric charge $\Omega$; Coulombic force is attractive for $\Omega Q < 0$, and repulsive for $\Omega Q > 0$. The radial force field breaks the symmetry of the string loop translation along the $z$ axis, and the string loop dynamics can not be regular even in the flat spacetime.

Using the condition $H_{\rm string}=0$, we find the effective potential (boundary function) $V_{\rm eff}(x,z)$ for a string loop electrically interacting with the point charge $Q$ in flat spacetime in the form ($y$ coordinate can be suppressed by fixing at $y=0$ then $r=\sqrt{x^2 + z^2}$)
\beq \label{energia}
V_{\rm eff}(x,z) =  \tens x + \frac{J^2}{x} +  \frac{\sqrt{2} \omega J Q}{r}.
\eeq
The case $\omega=-1$ corresponds to the configuration of opposite charges $Q$ and $\Omega$, the $\omega=0$ case to uncharged string loop, and the case $\omega=1$ corresponds to configurations with the same sign of the electric charges. We give examples of the effective potential function $V_{\rm eff}(x,z)$ in Fig. \ref{RNflat} and \ref{RN3D}.

In Fig.~\ref{RNflat} the left graph represents the string loop effective potential function $V_{\rm eff}(x,z=0)$ as section at the equatorial plane. In the $x$-direction, we have one minimum of the effective potential, depending on values of $J$ an $\omega$. In middle graph is plotted the string loop effective potential function $V_{\rm eff}(x=x_{\rm min},z)$ as section at its equatorial minimum. The stationary points of the $V_{\rm eff}(x,z)$ function are located in the equatorial plane, $z=0$, only; in the $z$-direction, for $\omega=-1$ we have minima, for $\omega=0$ we have constant behaviour in $z$ direction, and for $\omega=1$ we have saddle point. The behaviour of the effective potential along the vertical $z$ axis $V_{\rm eff}(x=x_{0},z)$, as section at $x_0=2$ (right Fig.), is also plotted for all three limiting values of $\omega$ parameter.

For visualizing the regions where the string loop motion is possible, we demonstrate in Fig.~\ref{RN3D}. the $E={\rm const.}$ sections of the effective potential full 2D function $V_{\rm eff}(x,z)$ for both $x$ and $z$ coordinates.
In the left, picture we give the $V_{\rm eff}$ profile for $\omega=-1$, $E=3.5$ case. The electrostatic interaction between the black hole and string loop charges is attractive and the string loop is trapped in closed area (light grey) - string loop is located in effective potential "lake".
In the middle figure, the case $\omega=0$, $E=4.5$ is presented. In this case, trapped motion in $x$ axis is observed, and there is no motion in vertical $z$ axis, if we consider zero initial velocity in $z$ direction, the effective potential (\ref{energia}) does not depend on $z$ in absence of the electric interaction between the black hole and string loop.
In the right picture, we shown the case $\omega=1$, $E=5$. The string loop is allowed to oscillate in a limited $x$ interval, while, due to the electrostatic repulsion between the black hole and string loop charges, the string loop is escaping along the vertical $z$ axis. Depending on the initial position, the string loop can move in the upper or lower half spaces and it can never cross the equatorial plane.

Coming from Fig.~\ref{RN3D}, we draw in Fig.~\ref{FlatTra} the trajectories of the string loop within their energy boundaries for the same values of the parameters $J$, $Q$ and $\omega$. We can conclude that in the case of opposite (attractive) charges ($\omega=-1$), electric attraction resists the string loop to escape to infinity. In the absence of electric interaction ($\omega=0$), there is no force in vertical direction and the trajectory of the string loop is always on the plane parallel to the $z$ plane. In the repulsively charged case ($\omega=1$), the electric repulsive potential barrier pushes the string loop away from the center.

%%%%%%%%%%%%%%%%%%%%%%%%%%%%%%%%%%%%%%%%%%%%%%%%%%%%%%%%%%%%%%%%%%%%%%%%%%%%%%%%%
\subsection{Charged string loops in Reissner-Nordstr\"{o}m background \label{loopRN}}

For string loop motion in the Reissner Nordstr\"{o}m background, the general form of the Hamiltonian (\ref{HamHam}) reduces to

\bea
H &=& \frac{1}{2}f(r) P_{r}^2+\frac{1}{2r^2} P_{\theta}^2+\frac{1}{2}\left (\frac{J^2}{r\sin\theta} + r\sin\theta\right)^2 \nn\\
&& -\frac{1}{2f(r)}\left(E-\frac{\Omega Q}{r}\right)^2.
\eea

As the whole axisymmetric string loop can be represented by a single point that can be characterized by a coordinate $y=0$ (see e.g. \cite{Stu-Kol:2012:JCAP:}), we can introduce the effective potential for charged string loop in the form
\beq \label{VeffRN}
V_{\rm eff}(x,z;Q,J,\omega)=\sqrt{1-\frac{2}{r}+\frac{Q^2}{r^2}}\left(\mu x + \frac{J^2}{x}\right) + \frac{\sqrt{2}\omega J Q}{r},
\eeq
where $r$ is radial distance $r^{2}=x^{2}+z^{2}$ and parameters $J,\omega$ were already introduced and explained in Eq. (\ref{JOdef}).  We put for simplicity $M=1$ (expressing $r$ in units of mass parameter). The effective potential $V_{\rm eff}(x,z)$ is not defined in the dynamical region, between the inner and outer RN black hole horizons, where $f(r) < 0$ \cite{Stu-Kol:2012:JCAP:}. For the black hole spacetimes, $Q\leq{1}$, we will consider string loop motion in the region above the outer horizon, $r > r_{+}$, see Eq. (\ref{RNhorizon}). For the RN naked singularity spacetimes, $Q>1$, the dynamical region ceases to exist, and $V_{\rm eff}(x,z)$ is defined for any $r>0$.

First we need to explore asymptotic behaviour of the effective potential (\ref{VeffRN}). Reissner Nordstr\"{o}m spacetime is asymptotically flat, hence in the x-direction there is
\beq
V_{\rm eff}(x \rightarrow \infty,Q,J,\omega)\rightarrow +\infty,
\eeq
and in the z-direction we obtain
\beq\label{Ey}
V_{\rm eff}(z \rightarrow \infty,Q,J,\omega)\rightarrow x + \frac{J^2}{x}=V_{\rm eff(flat)},
\eeq
for details see~\cite{Kol-Stu:2010:PHYSR4:}.

Here we consider all possible types of the charged string loop motion around the RN black holes as well as the RN naked singularities. The case of the charged string loop motion in the field of RN black holes and naked singularities is included in the related study of behavior of string loops in the braneworld spherically symmetric black holes studied in \cite{Stu-Kol:2012:JCAP:}, that where it was demonstrated that the string loop can oscillate in the closed area, fall down into the black hole, or escape to infinity in the vertical direction, while oscillating in the $x$-direction. Exploring the effective potential the type of string loop motion can be estimated.

\begin{figure*}
\includegraphics[width=\hsize]{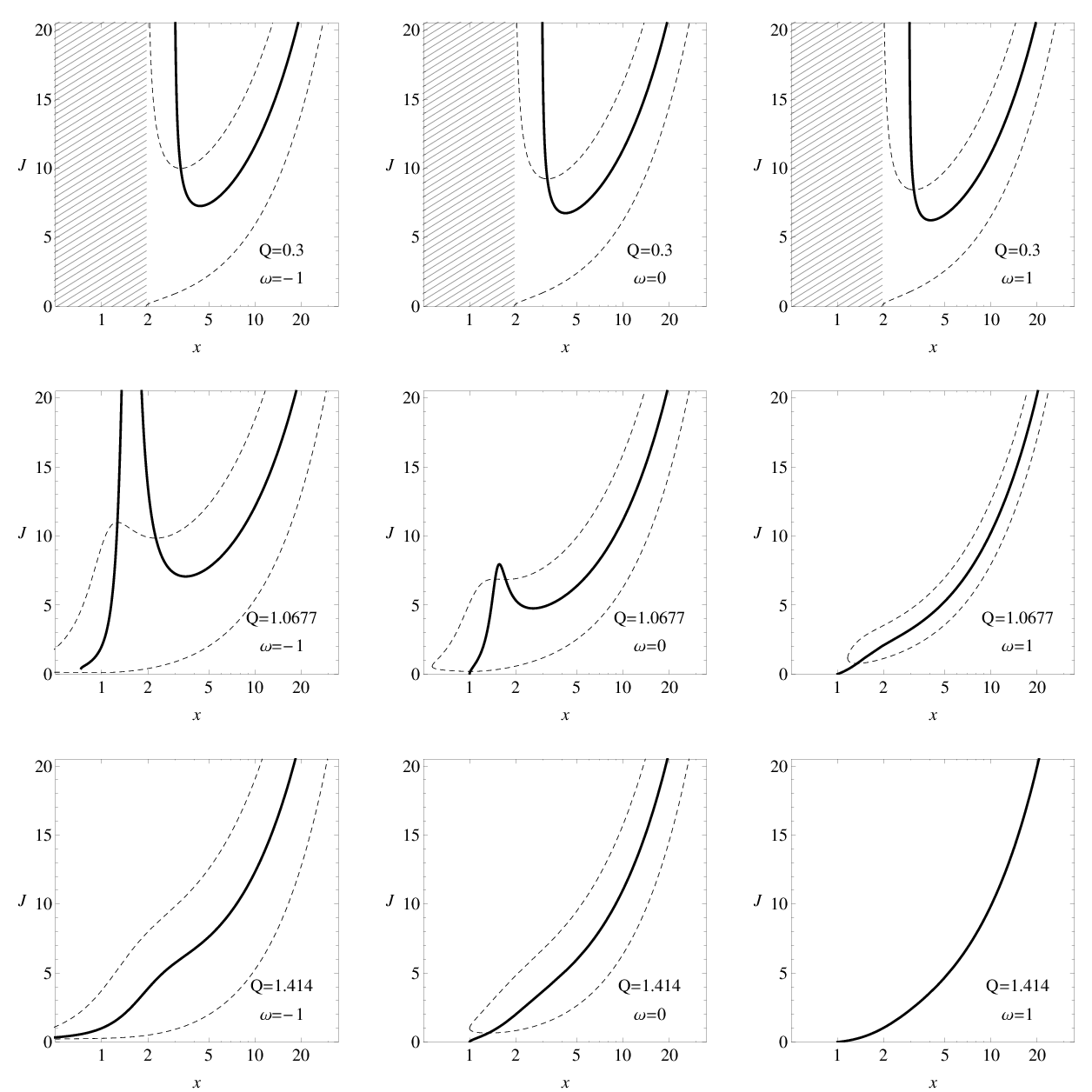}
\caption{\label{jex} The behavior of string loop's angular momentum parameter corresponding to effective potential's equatorial extreme (thick black) along with angular momentum parameters $J_{\rm L1}$ and $J_{\rm L2}$ defining the trapped motion boundaries (dashed). If the $J_{\rm ext}$'s profile is within the shaded lines string loop's motion is always in some toroidal space otherwise it escapes to infinity.}
\end{figure*}

Stationary points of the 2D effective potential function $V_{\rm eff}(x,z)$ are given by Eq. (\ref{extr}). The stationary points can be found in the equatorial plane, $z=0$, and their $x$ coordinate is given by the relation
\bea \label{deff}
\frac{H^2}{x}\left(\mu-\frac{J^2}{x^2}\right)+(x-Q^2)\left(\frac{J^2}{x^2}+\mu\right) && \nn\\
-\frac{\sqrt{2}JQ\omega}{x}H&=&0.
\eea
From Eq. (\ref{deff}) one can easily find the corresponding condition for string loop angular momentum parameter $J$
\beq
 J = J_{\rm ext},
\eeq
where
\beq \label{j}
J_{\rm ext}\equiv\frac{-Q H \omega x \pm x \sqrt{2P(x-1)x+Q^2H^{2}\omega^2}}{\sqrt{2} P}.
\eeq
Here we have used the following notations:
\bea
P(x,Q)&=&2Q^2-3x+x^2, \nn\\
H(x,Q)&=&\sqrt{Q^2-2x+x^2}.
\eea
The $J_{\rm ext}(r;\omega,Q)$ function determines both stable and unstable stationary positions of the string loop. Radial profiles of $J_{\rm ext}(r;\omega,Q)$ function are plotted in Fig.~\ref{jex}. for various combinations of $\omega$ and $Q$ charges.

Depending on the black hole charge $Q$ and string loop charge parameter $\omega$, there can exist one, two or three stationary points of the effective potential in the equatorial plane. The sign of $\frac{dJ_{\rm ext}}{dx}$ defines type of the extrema, the positive derivation term, $\frac{dJ_{\rm ext}}{dx}>0$, determines the effective potential equatorial minima while negative derivative, $\frac{dJ_{\rm ext}}{dx}<0$, determines the maxima. The extremal point of the $J_{\rm ext}$ function, given by $\frac{dJ_{\rm ext}}{dx}=0$, defines the  innermost stable string loop position.

Exploration of the effective potential $V_{\rm eff}$ allows us to determine all possible types of the string loop trajectories in the RN backgrounds. In addition to the effective potential extrema, given by the $J_{\rm ext}$ function, we need to explore when the string loop can escape to infinity along the $z$ axis. From Eq.~(\ref{Ey}) we know that the energetic boundary will be open in the $z$ direction for the string loop motion, if string energy $E$ satisfies the condition
\beq \label{Einf}
E > V_{\rm eff(flat)} = 2J.
\eeq
Therefore, condition $V_{\rm eff}(x,z;Q,\omega)<2J$ gives the boundaries of string loop's trapped motion. Solving the Eq. \ref{Einf} with respect to $J$, we find the regions where string loop's motion is trapped. We thus derive the loop trapping function in the form
\beq
J_{L1,2} = \frac{-Q x^3 H \pm x\sqrt{(Q^2\omega^2H-2(x^2-x)P)}}{\sqrt{2}P}.
\eeq

The trapped oscillatory string loop motion occurs under the circumstance given by the relations
\beq
J_{L1}(x;Q,\omega) >J> J_{L1}(x;Q,\omega).
\eeq
In Fig.~\ref{jex} we demonstrate the behavior of the extrema $J_{\rm ext}$ function, along with the boundary functions $J_{L1}$ and $J_{L2}$ giving trapped motion boundaries; if the string loop angular momentum is located within the dashed boundaries the string loop can never escape to infinity in the vertical direction at the corresponding fixed $x$.

%%%%%%%%%%%%%%%%%%%%%%%%% eff potential extrem - J_{ext} function
\begin{figure*}
\includegraphics[width=\hsize]{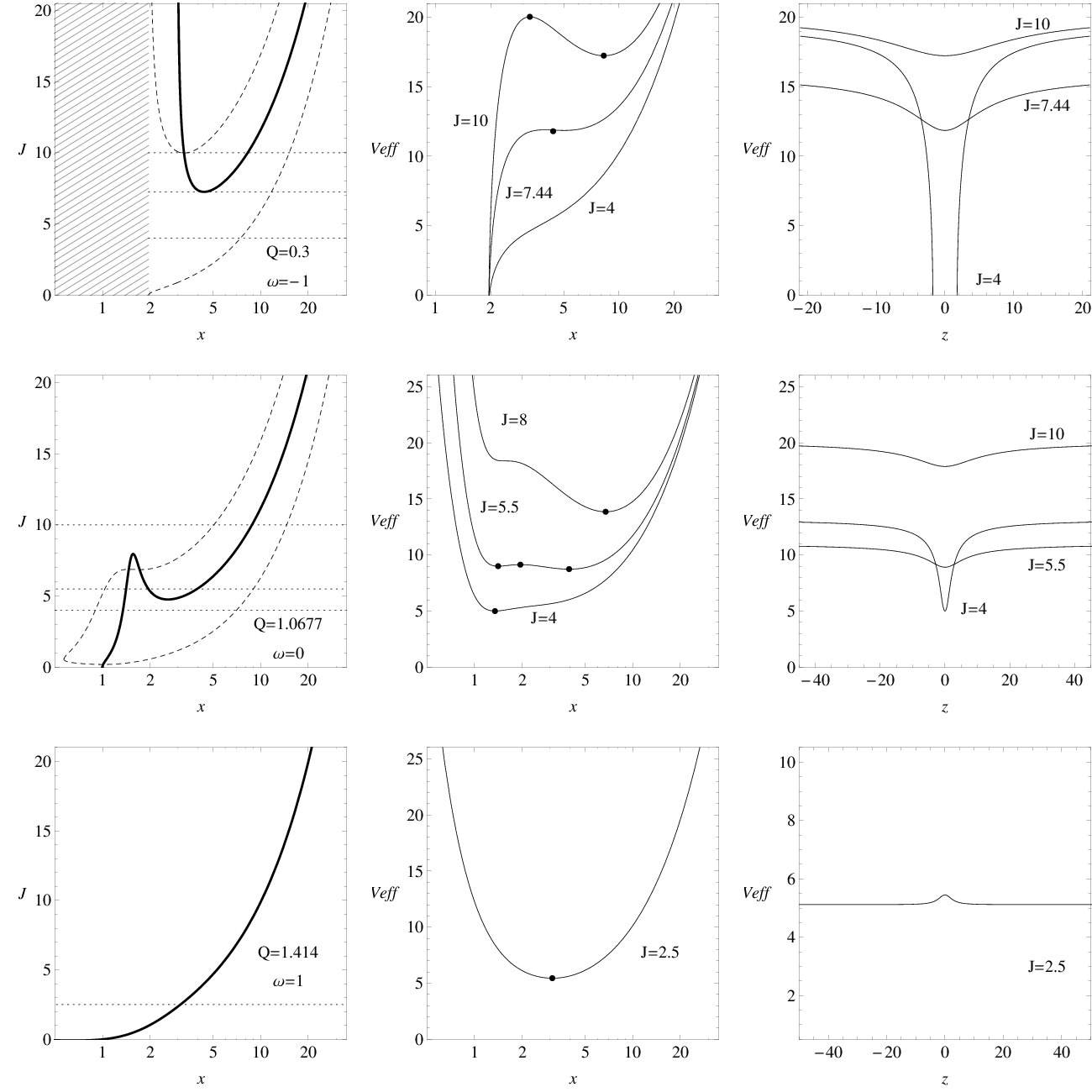}
\caption{\label{jeff} The behavior of $J_{\rm ext}$ function and boundary angular momentum parameter functions $J_{\rm L1}$, $J_{\rm L2}$ and the corresponding effective potentials. Dots denote the extreme points of effective potential $V_{\rm eff}$. Shaded region stands for area below black hole horizon.}
\end{figure*}
%%%%%%%%%%%%%%%%%%%%%%%%%

%%%%%%%%%%%%%%%%%%%%%%%%%    string loop trajectories

% different shapes of V_{eff} - different types of string loop motion

\begin{figure*}
\subfigure[ \,\, $J=10$]{\includegraphics[width=0.32\hsize]{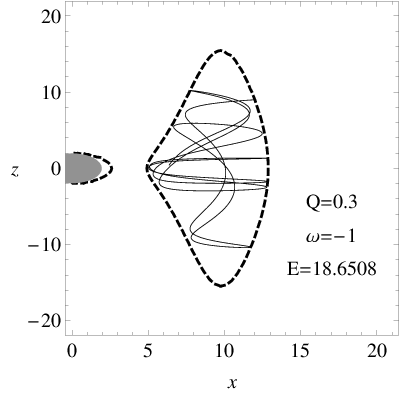}}
\subfigure[ \,\, $J=7.44$]{\includegraphics[width=0.32\hsize]{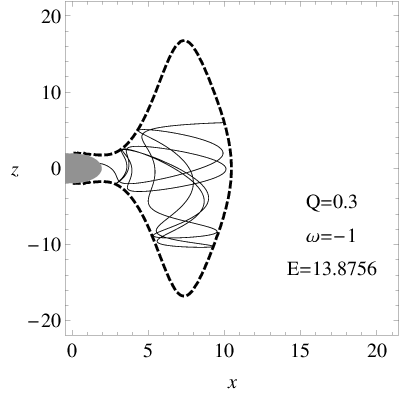}}
\subfigure[ \,\, $J=4$]{\includegraphics[width=0.32\hsize]{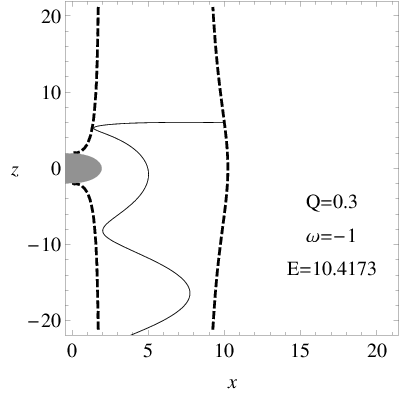}}
\subfigure[ \,\, $J=10$]{\includegraphics[width=0.32\hsize]{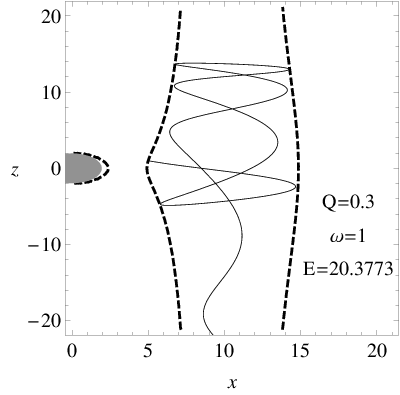}}
\subfigure[ \,\, $J=5.5$]{\includegraphics[width=0.32\hsize]{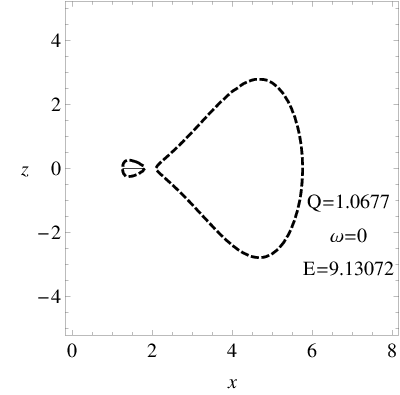}}
\subfigure[ \,\, $J=6.5$]{\includegraphics[width=0.32\hsize]{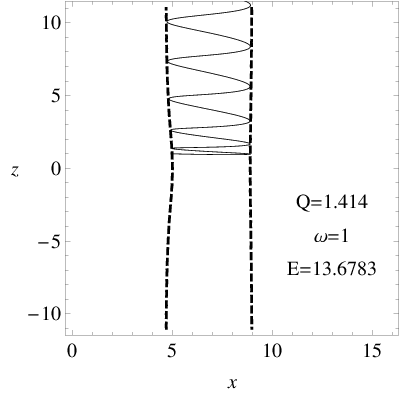}}
\caption{\label{Traj} (prepsat) Trajectories of the string loop and energy boundaries of their motion. Four various types of motion possible: trapped in some "lake"-like region (($a$), ($e$)), capture by black hole ($b$), collapse or escape to infinity ($c$), escape to infinity (($d$), ($f$)).}
\end{figure*}

In Fig.~\ref{jeff} we use the diagonal pictures of Fig.~\ref{jex} and construct the corresponding effective potential along $x$-axis and $z$-axis at $x_{\rm min}$ given for the chosen values of $J$, where $x_{\rm min}$ is position of the effective potential minima.
In Fig. \ref{Traj} we give some typical trajectories of the string loop motion.
In the top row of the Fig.~\ref{jeff}, we consider $Q=0.3$, $\omega=-1$ case. First we take $J=10$ as it is crossing the $J_{\rm ext}$ profile at two points (Fig~\ref{jeff}). In this case, there is one stable and one unstable equilibrium position.

In Fig.~\ref{Traj}(a) we show trajectories of string loop's motion cross-section along with the boundary energy profiles - we observe that the motion is trapped in the closed region. Then we consider $J=7.44$, as this value of $J$ is touching the extremal point of the function $J_{\rm ext}$(Fig~\ref{jeff}), corresponding to the innermost stable equilibrium position (\ISSP) - any small deviation from this position causes the string to collapse to the black hole. String loop's trajectory for this case is given in Fig.~\ref{Traj}(b); we can conclude that the motion is finite in the $z$-direction and the energy boundary profile is open to the black hole, and the string finally falls down to the black hole. And last case of $Q=0.3$, $\omega=-1$ configuration, we take $J=4$ value as it is not crossing the $J_{\rm ext}$ profile at all. In this case, there is no possible trapped motion and the string loop has to escape to infinity in the vertical direction (Fig.~\ref{Traj}(c)). Another possible string loop trajectory around the black hole is given for $Q=0.3$, $\omega=1$, $J=10$ case in Fig.~\ref{Traj}(d), with string escaping to infinity.

Medium line elements of Fig.~\ref{jeff} represent the naked singularity $Q=1.0677$, $\omega=0$ case. The most distinctive behavior of the effective potential is given by the presence of two minima for $J=5.5$. This indicates that the string loop has in the $x$-direction two stable positions around the naked singularity, and escape along vertical direction is impossible for sufficiently low string loop energy. The trajectory of the string loop for this type of motion is given in Fig.~\ref{Traj}(e).
This type of energy boundary profile corresponds to trapped motion - the trapped motion can take place in one of two possible closed toroidal spaces around the RN naked singularity. At the bottom line on Fig.~\ref{jeff} we consider $Q=1.414$, $\omega=1$ situation. There is one minimum of the potential well corresponding to stable equilibrium position. String loop's motion in $x$-direction is limited. There is also small potential barrier resisting the string loop to cross the $z=0$ equatorial plane. String loop escapes to infinity loosing oscillatory energy in the $x$ direction (Fig.~\ref{Traj}(f))~\cite{Kol-Stu:2010:PHYSR4:, Lar:1994:CLAQG:, Jac-Sot:2009:PHYSR4:}.

%%%%%%%%%%%%%%%%%%%%%%%%%    string loop ISSP vs. particle ISCO

\begin{figure*}
\includegraphics[width=\hsize]{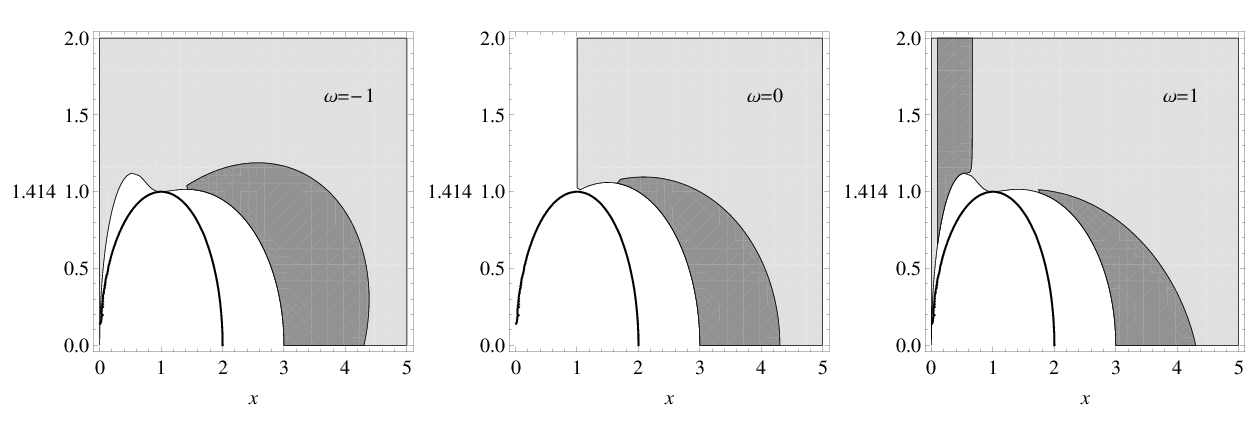}
\caption{\label{QX} Local extrema position $x$ and type (maxima/minima) of the effective potential $V_{\rm eff}(r,\theta=\pi/2)$ for different values of RN charge parameter $Q$ and for all three considered values of $\omega$ parameter. Thick black line corresponds to event horizon and restrict the dynamical region. 
Darker grey colour denotes region where maxima of the $V_{\rm eff}$ function can exist, while lighter grey colour denotes region where minima can exist. Only in the lighter grey areas can exist stable string loop position - the boundary between darker/light areas act as innermost stable string loop position. 
In white areas above RN black hole horizon there are no extrema point of $V_{\rm eff}$ function.}
\end{figure*}

Effective potential $V_{\rm eff}(x,z=0)$ has real extrema only for real values of the extreme angular momentum function $J_{\rm ext}$ given by expression (\ref{j}).  We thus find the condition relating the limiting values of RN charge parameter $Q$ and the string loop charge parameter $\omega$  in the form
\beq
 \omega^2 = \omega^2_{\rm crit}(x,Q) \equiv - \frac{Q^2(2 Q^2 -3x +x^2)(x-1)}{Q^2 x^2 (Q^2 -2x +x^2)},
\eeq
along with the condition $Q^2 -2x +x^2 \geq 0$. This allow us to distinguish regions with different string loop effective potential behavior for any central charge $Q$, as demonstrated in Fig.~\ref{QX}, where the region around a black hole or naked singularity is separated into regions where effective potential has minima (light grey region), or maxima (grey region), or there are no extrema at all (white region). The line dividing grey and light grey regions gives the location of innermost stable equilibrium position (\ISSP).

\begin{figure*}
\includegraphics[width=\hsize]{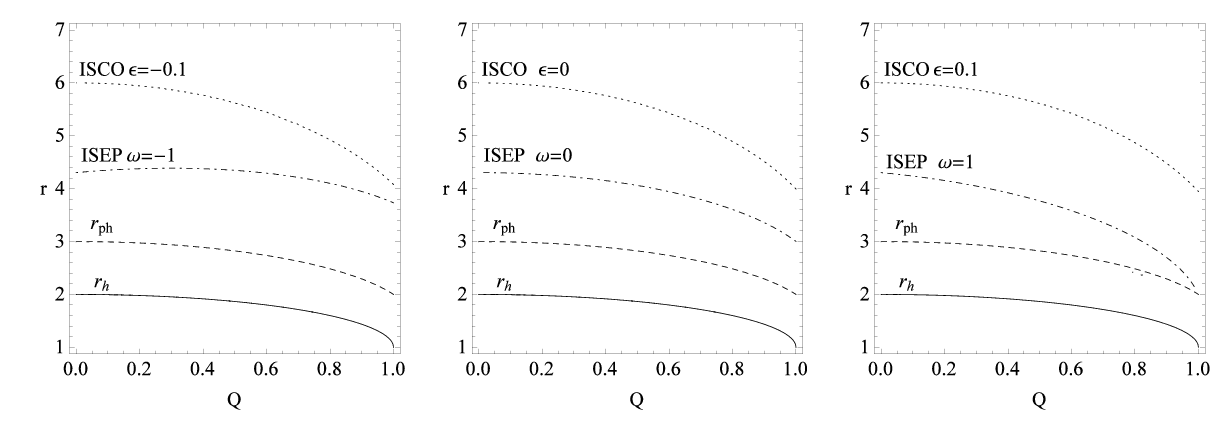}
\caption{\label{figISCO} ISEP for string loop and ISCO for charged particle with respect to black hole charge $Q$. ISEP for string are given for attracting $\omega=-1$, not interacting $\omega=0$ and repulsing $\omega=1$ interaction types between string and charged black hole. ISCO for charged particles are defined for three different charge to mass ratio $\epsilon$ values.}
\end{figure*}

It is useful to compare the \ISSP \ for charged string loops with its particle equivalent - the charged particle innermost stable circular orbit (ISCO), in the same RN black hole background \cite{Pug-Que-Ruf:2011:PHYSR4:, Pug-Que-Ruf:2017:EPJC:}. In Fig.~\ref{figISCO}, such comparison is given for all three (positive, neutral, negative) variants of the charged test object. As it can be seen, the charged string loop \ISSP \ is always located between the photon orbit and the charged particle ISCO in the RN spacetime, revealing true about the string loop real nature.

%%%%%%%%%%%%%%%%%%%%%%%%%%%%%%%%%%%%%%%%%%%%%%%%%%%%%%%%%%%%%%%%%%%%%%%%%%%%%%%%%

\section{Quasi-periodic oscillations of string loops}

%%%%%%%%%%%%%%%%%%%%%%%%%%%%%%%%%%%%%%%%%%%%%%%%%%%%%%%%%%%%%%%%%%%%%%%%%%%%%%%%%

The quasi-periodic oscillatory motion of the string loops trapped in a toroidal space (or in "lake") around the minima of the effective potential $V_{\rm eff}(x,z)$ function could be used to interpret interesting astrophysical phenomenon - high-frequency quasi-periodic oscillations (HF QPOs). Most of compact X-ray binaries that contain a black hole or a neutron star demonstrate quasi-periodic variability of the X-ray flux in the kHz frequency range. Some of these HF QPOs appear in pairs as upper and lower frequencies ($\nu_{\rm U}, \nu_{\rm L}$) and in Fourier spectra are observed twin peaks. Since the peaks of high frequencies are close to the orbital frequency of the marginally stable circular orbit representing the inner edge of Keplerian discs orbiting black holes (or neutron stars), the strong gravity effects must be relevant to interpret HF QPOs~\cite{Stu-Kol:2014:PHYSR4:}. So far, many models have been proposed to explain HF QPOs in black hole binaries: the relativistic precession model, the warped disc model, resonance model~\cite{Tor-etal:2005:ASTRA:, Mot:2016:ANOTES:, Stu-Kol:2016:APJ:,Stu-Kol:2015:MNRAS:,Stu-Kol:2016:ASTRA:}. Usually, Keplerian orbital and epicyclic (radial and latitudinal) frequencies of geodetical circular motion are assumed in models explaining the HF QPOs in both black hole and neutron star systems~\cite{Stu-Kot-Tor:2013:ASTRA:}. However, neither of these models is able to explain the HF QPOs in all microquasars~\cite{Tor-etal:2011:ASTRA:}. On the other hand, there is possibility of the relevance of string loop's oscillations, characterized by their radial and vertical (latitudinal) frequencies that are comparable to the epicyclic geodetical frequencies, but slightly different, enabling thus some corrections to the predictions of the models based on the geodetical epicyclic frequencies. Of course, the frequencies of string loops oscillations in physical units have to be related to distant observers.

Let have string loop located in the equatorial plane and at a effective potential $V_{\rm eff}(r,\theta)$ minimum ($r=r_0, \theta=\theta_0=\pi/2$). Slight displacement from minima position $r=r_0 + \delta r, \theta=\theta_0 + \delta \theta$, causes small string loop oscillations around the stable equilibrium positions, determined by the equations of harmonic oscillations
\beq
 \ddot{ \delta r} + {\omega_r}^{2}\delta r=0, \quad  \ddot{ \delta \theta} + {\omega_\theta}^{2}\delta \theta=0,
\eeq
where locally measured frequencies of the oscillatory motion are given by
\beq
 \omega_r^{2}=\frac{\partial ^2 V_{\rm eff}}{\partial r^2}, \quad   \omega_\theta^{2}=\frac{1}{r^2 f(r)}\frac{\partial ^2 V_{\rm eff}}{\partial \theta^2}.
\eeq

Local observers, at the position of the string loop in the RN spacetime measure angular frequencies

\bea\label{fq2}
\omega_r^{2}&=&\frac{2 \left(J_{\rm ext}^2 \left(6 Q^2+(r-6) r\right)+\sqrt{2} J_{\rm ext} Q r^2 \omega +Q^2 r^2\right)}{r^5},\nn\\
\omega_\theta^{2}&=&\frac{(J_{\rm ext}-r) (J_{\rm ext}+r)}{r^3}.
\eea

where $J_{\rm ext}(r)$ gives $J$ parameter for equatorial minima. Frequencies measured by static observers at infinity $\Omega$ are related to the locally measured frequencies (\ref{fq2}) by the gravitational redshift transformation
\beq
\Omega_{(r,\theta)}=\frac{\d X_{(r,\theta)}}{\d t}= \frac{\d X_{(r,\theta)}}{\d \zeta} \frac{\d \zeta}{\d t} = \frac{\omega_{(r,\theta)}}{E}\ f(r),
\eeq
here $E=E(r_0,\theta_0)$ is the energy of the string loop on its minima position and $f(r)$ is the characteristic lapse function of the RN metric (\ref{SfSymMetrika}). The frequencies for observers at infinity $\Omega$, have to be multiplied by the factor $c^3/GM$ to be expressed in the standard physical units
\beq
\nu_{(r,\theta)}=\frac{1}{2\pi}\frac{c^3}{G M}\ \Omega_{(r,\theta)}.
\eeq

\begin{figure*}
\includegraphics[width=\hsize]{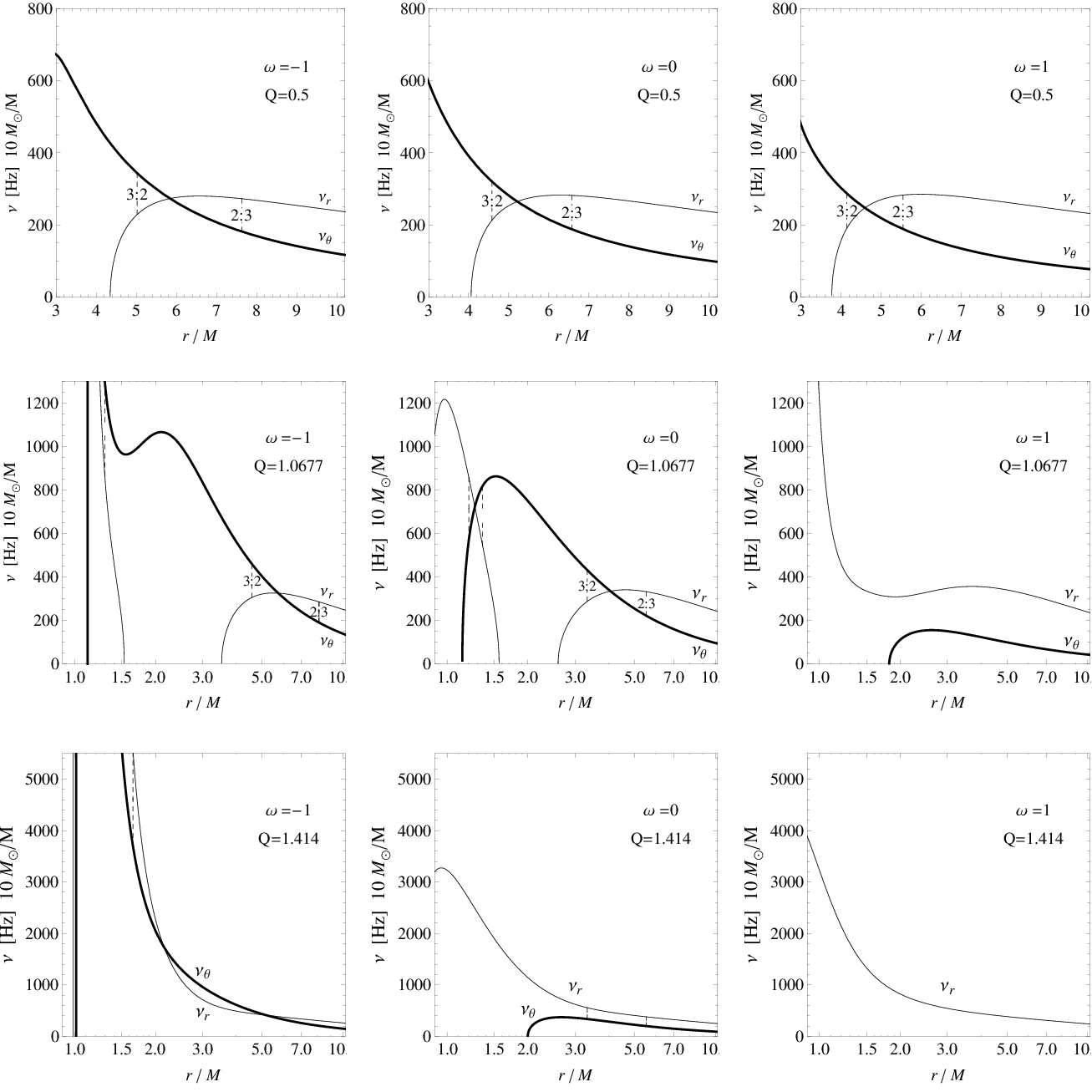}
\caption{\label{qx} Radial profiles of radial and vertical fundamental frequencies $\nu_r$, $\nu_{\theta}$ measured by distant observers, determined for harmonic string loop oscillations around stable equatorial equilibrium states. String loops are considered around Reissner-Nordstr\"{o}m black hole of 10 solar masses. Radii of $3:2$ and $2:3$ resonances are given as dashed lines.}
\end{figure*}

We focus our attention to resonance frequencies with ratio 3:2 observed in X-ray data from GRO 1655-40, XTE 1550-564, and GRS 1915+105 that require the string loop frequencies corresponding to twin peaks appears in the $\nu_{r}:\nu_{\theta}=3:2$ or $\nu_{\theta}:\nu_{r}=3:2$ ratio - see Tab.~\ref{tab1}. We explore the displacement of resonance frequencies with respect to black hole's charge $Q$ and string loop $\omega$ parameter. In Fig.~\ref{qx} we illustrate the radial coordinates of equilibrium positions where $\nu_{r}:\nu_{\theta}=3:2$ or $\nu_{\theta}:\nu_{r}=3:2$ ratios appear in dependence on parameters $Q$ and $\omega$. On the top row of Fig.~\ref{qx} we present $Q=0.5$ case. As it is seen, with changing from strong electric attraction, $\omega=-1$, to absence of interaction, $\omega=0$, and finally to strong electric repulsion, $\omega=1$, the position of resonance frequencies tend to come closer to the black hole. For the naked singularity case of $Q=1.0677$ on the middle row of Fig.~\ref{qx}, in the similar step of changes of $\omega$, we observe different scenario from $Q=0.5$ situation. Here, for $\omega=-1$ case three equilibrium points satisfying the 3:2 ratio condition for $\nu_r$ and $\nu_\theta$ occur. Further, in $\omega=0$ case, the resonance frequencies occur at four positions. Finally, when $\omega=1$, the resonance frequencies does not appear at all. Bottom line elements on Fig.~\ref{qx} represent the $Q=1.414$ naked singularity case. In this scenario, the resonance 3:2 frequencies appear at two locations for the $\omega=-1$ case, while for the $\omega=0$ case they occur only at one position, and in the $\omega=1$ case vertical frequencies disappear, the string loops are unstable relative to vertical perturbations.

\begin{table*}
\begin{center}
\begin{tabular}{ c c c c }
\hline  Source & GRO 1655-40 & XTE 1550-564 & GRS 1915+105   \\[0.5ex]
\hline
\hline
$\nu_U$ [Hz] & 447\ ---\ 453& 273\ ---\ 279& 165\ ---\ 171 \\[1ex]
$\nu_L$ [Hz] & 295\ ---\ 305& 179\ ---\ 189& 108\ ---\ 118 \\[1ex]
$M/M_{\odot}$ & 6.03\ ---\ 6.57& 8.5\ ---\ 9.7& 9.6\ ---\ 18.4 \\[0.5ex]
\hline
\end{tabular}
\caption{\label{tab1} Observed twin HF QPOs data for three microquasars, and the restrictions on mass of black holes located in them, based on independent measurements on the HF QPO measurements.}
\end{center}
\end{table*}

\begin{figure*}
\subfigure[ \,\, $Q=0$]{\includegraphics[width=0.32\hsize]{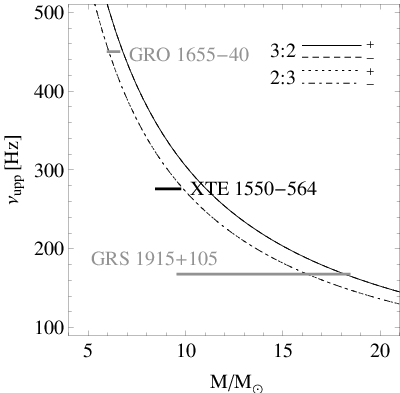}}
\subfigure[ \,\, $Q=0.5$]{\includegraphics[width=0.32\hsize]{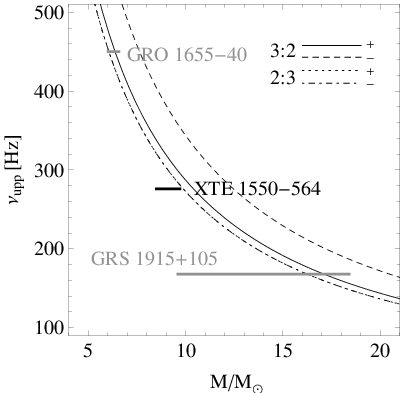}}
\subfigure[ \,\, $Q=0.8$]{\includegraphics[width=0.32\hsize]{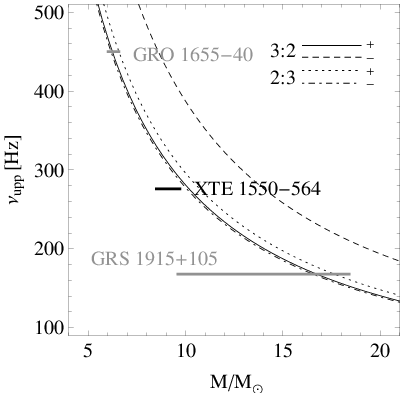}}
\caption{\label{rezM} The upper frequency $\nu_U$ of string loop oscillations at 3:2 or 2:3 resonant radii, calculated in the framework of the string loop model with maximal range string loop parameter $\omega$ as a function of black hole mass for several values of the black hole charge $Q=0, 0.5, 0.8$. }
\end{figure*}

For fixed black hole charge $Q$ and fixed string loop charge parameter $\omega$, upper frequency of the twin HF QPOs can be given as a function of black hole's mass $M$. If the black hole mass is restricted by separated observations, as is commonly the case, we obtain some restrictions on the string loop resonant oscillations model, as illustrated in Fig.~\ref{rezM}. Here, the situation is demonstrated for several values of black hole's charge $Q$ and limits on the black hole mass as given in Tab.~\ref{tab1}. We can see that for the Schwarzschild black hole ($Q=0$), the string loop model can explain only the HF QPOs in GRS 1915+105. Introducing black hole charge $Q$ and parameter $\omega$, the string loop resonant oscillations model widens the area of its applicability. For $\omega=1$  and $Q=0.5, 0.8$ case, the model fully describes observed values from GRO 1655-40 source. It contains the whole range of expected mass range from Tab.~\ref{tab1}. Nevertheless, the string loop resonant oscillation model in Reissner-Nordstr\"{o}m background can not explain the observed values from XTE 1550-564 source. For any value of $\omega$ parameter and for any low values of black hole charge $Q$, the string loop model can not fit observed mass range for the XTE 1550-564 source and an additional influence of the black hole rotation has to be expected.

Moreover, in Fig. \ref{rezM} we can clearly see that the predicted value of the black hole mass is increasing with the black hole charge $Q$ increase. It will become harder and harder to fit the observed HF QPOs as the $Q$ parameter increases, hence we can conclude that introducing new parameter $Q$ into the string loop HF QPOs model is not successfully efficient in explaining the observed HF QPOs in microquasars, and inclusion of the black hole spin that can be sufficiently efficient as demonstrated in \cite{Stu-Kol:2014:PHYSR4:} is necessary.

%%%%%%%%%%%%%%%%%%%%%%%%%%%%%%%%%%%%%%%%%%%%%%%%%%%%%%%%%%%%%%%%%%%%%%%%%%%%%%%%%

\section{String loop acceleration and asymptotical ejection speed}

%%%%%%%%%%%%%%%%%%%%%%%%%%%%%%%%%%%%%%%%%%%%%%%%%%%%%%%%%%%%%%%%%%%%%%%%%%%%%%%%%

From the astrophysical perspective, one of the most relevant applications of the axisymmetric string loop motion is the possibility of strong acceleration of the linear string motion due to the transmutation process in the strong gravity field of immensely compact objects that arises due to the chaotic character of the string loop motion and could well simulate acceleration of relativistic jets in Active Galactic Nuclei (AGN) and microquasars~\cite{ Jac-Sot:2009:PHYSR4:, Stu-Kol:2012:PHYSR4:, Kol-Stu:2013:PHYSR4:}. Since the RN spacetime is asymptotically flat, we have to examine the linear string loop motion in the flat spacetime; as the Columbian electric field disappears asymptotically at the RN spacetime, this approximation is sufficient to understood the results of the acceleration process. The energy of string loop (\ref{energia}) in the Cartesian coordinates reads
\beq
E^{2}=\dot{z}^{2}+\dot{x}^{2}+\left(\frac{J^{2}}{x}+x\right)^{2}=E_{\rm z}^{2}+E_{\rm x}^{2},
\eeq
where dot denotes derivative with respect to the affine parameter $\zeta$. The energies related to the $x-$ and $z-$ directions are given by the relations
\beq
\label{ezex}
E_z^{2}=\dot{z}^{2}, \qquad E_x^{2}=\dot{x}^{2}+\left(\frac{J^{2}}{x}+x\right)^{2}=(x_i+x_o)^2=E_0^2,
\eeq
where $x_i$ ($x_o$) represents inner (outer) boundary of the oscillatory motion. The energy $E_0$ representing the internal energy of the string loop is minimal when the inner and outer radii coincide, leading to the relation
\beq
E_{\rm 0(min)}=2J
\eeq
that determines the minimal energy needed for escaping of the string loop to the infinity in the spacetimes related to black holes or naked singularities.

Clearly, $E_{\rm x}=E_0$ and $E_{\rm z}$ are constants of string motion in the flat spacetime and transmutation between energy modes are not possible there. However, in the vicinity of black holes, the kinetic energy of oscillating string can be transformed into the kinetic energy  of the translational linear motion.

The energy in the x-direction $E_0$ can be interpreted as an internal energy of the oscillating string, consisting from the potential and kinetic parts; only in the limiting case of $x_{\rm i}=x_{\rm o}$, the internal energy has zero kinetic component. The string internal energy can quite reasonably represent the rest energy of string moving in the $z$-direction in flat spacetime~\cite{Stu-Kol:2012:PHYSR4:}. The final Lorentz factor of the translational motion of an accelerated string loop as observed in asymptotically flat region of the Reissner-Nordstr\"{o}m spacetimes is, from (\ref{ezex}) defined by the relation~\cite{Jac-Sot:2009:PHYSR4:, Stu-Kol:2012:PHYSR4:}.
\beq
\gamma=\frac{E}{E_0}=\frac{E}{x_{\rm i}+x_{\rm o}},
\eeq
where $E$ is the total energy of the string loop moving with the internal energy $E_0$ in the $z$-direction, with the velocity corresponding to the Lorenz factor $\gamma$. Apparently, the maximal Lorentz factor of the transitional motion reads~\cite{ Stu-Kol:2012:PHYSR4:}
\beq
\gamma_{\rm max}=\frac{E}{2J} .
\eeq

\begin{figure*}
\includegraphics[width=\hsize]{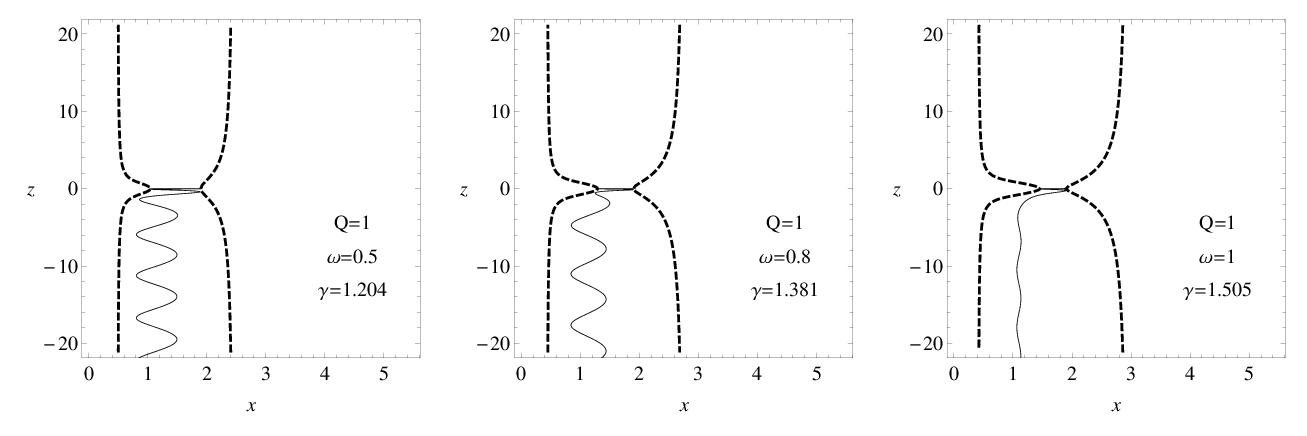}
\caption{\label{gamma} Escaping trajectories of string in flat spacetime with respect to string charge parameter $\omega$. Asymptotical Lorenz factor $\gamma$ of transmitted string loops is given for three values of $\omega$ parameter for the value of black hole charge $Q=1$. The Lorenz factor $\gamma$ is calculated for string loops with angular momentum $J=1.1$ and starting from initial position $x_0=1.9, z_0=0$.}
\end{figure*}

From this equation we can see that for observing ultra-relativistic acceleration of the string loop large ratio of the string energy $E$ versus its angular momentum parameter $J$ is needed. In Fig.~\ref{gamma} we illustrate asymptotic linear speed of transmitted string loops. We demonstrate the influence of the $\omega$ parameter on ejection speed of string loops for extremal black holes with charge $Q=1$, string angular momentum parameter $J=1.1$, starting from position $x_0=1.9, z_0=0$. Ejection speeds are expressed by the Lorentz factor ($\gamma=1/\sqrt{1-v_{\rm ejection}^2}$). As presented in Fig.~\ref{gamma}, for bigger values of $\omega$ we observe greater values of ejection speed. This can be explained due to repulsion from the center of the acceleration of the string loop. Nevertheless, the observed ejection speeds are not so highly relativistic as they are in the Kerr naked singularity spacetimes~\cite{Kol-Stu:2013:PHYSR4:}.

\begin{figure*}
\includegraphics[width=\hsize]{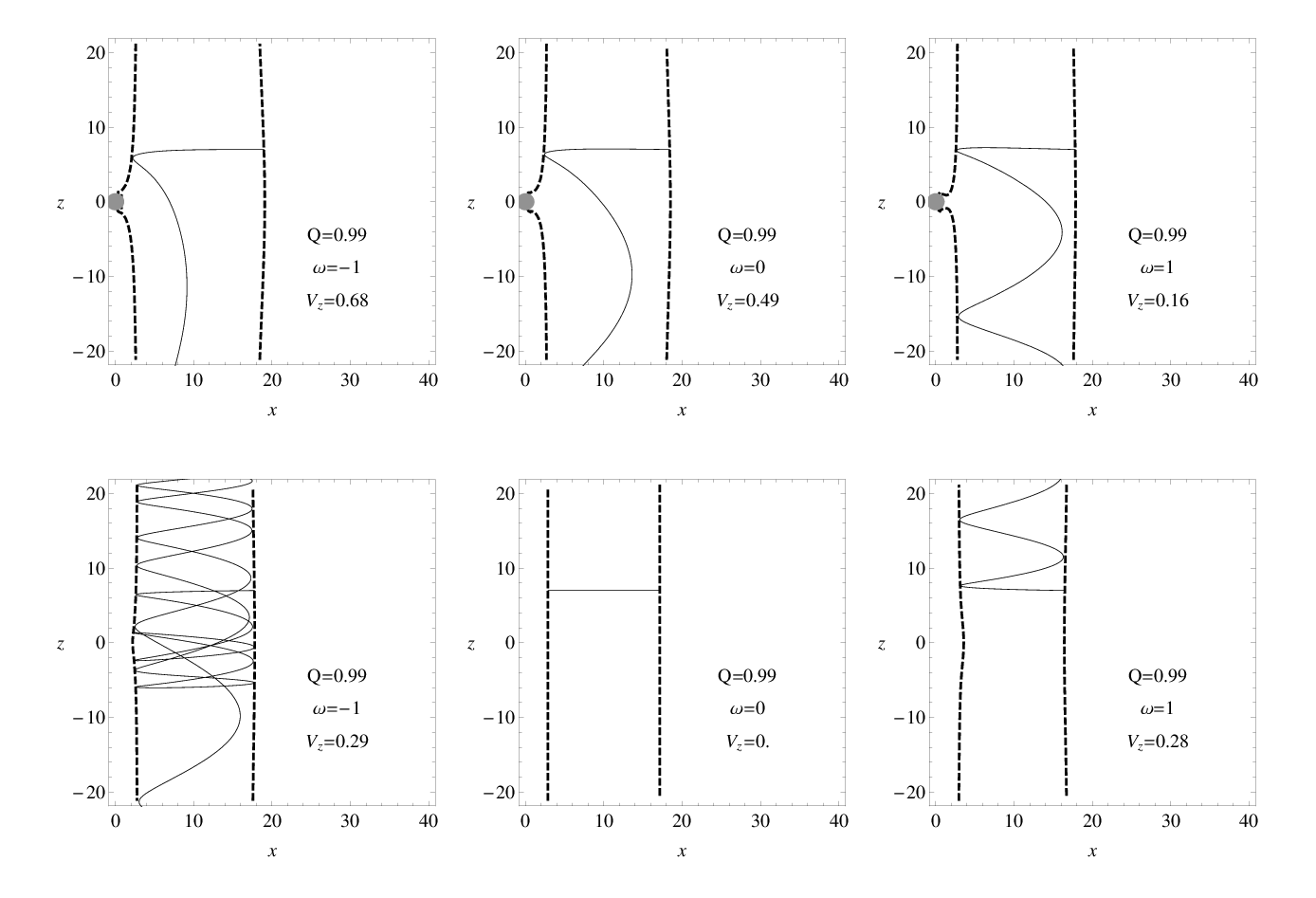}
\caption{\label{RN12} Escaping trajectories of string loop in RN(top line elements) and flat spacetime(bottom line elements). Asymptotical speed of transmitted string loops given for three values of $\omega$ parameter for charged black hole $Q$ and flat spacetime. The asymptotical ejection speed of string loop $V_{z}$ is calculated for the values of angular momentum $J=7$, energy $E=20$ and initial position $z_0=7$. $x_0$ is found from the expressions of energy in RN~(\ref{VeffRN}) and flat spacetime~(\ref{energia}).}
\end{figure*}

In Fig.~\ref{RN12} we give escaping trajectories of the transmitted string loops in the RN background and flat spacetime for attracting, $\omega=-1$, not interacting, $\omega=0$, and repulsing, $\omega=1$, types of the string loop interaction with the charged black hole and central point charge in flat spacetime. It is expected to observe bigger ejection speed in repulsing case($\omega=1$) than attracting one($\omega=-1$). However, surprisingly the string loop acceleration is higher in $\omega=-1$ case than $\omega=1$ case. This can be explained by studying their trajectories within the energy boundaries. In the $\omega=-1$ case, due to attraction by the black hole, the string loops enters deeper in a black hole's potential well and the transition effect of oscillating energy to escaping translational energy in the $z$-direction becomes more effective. 
Due to the chaotic nature of string loop dynamics, we can expect completely different set of velocities for different set of initial conditions.

%%%%%%%%%%%%%%%%%%%%%%%%%%%%%%%%%%%%%%%%%%%%%%%%%%%%%%%%%%%%%%%%%%%%%%%%%%%%%%%%%

\section{Conclusion}

%%%%%%%%%%%%%%%%%%%%%%%%%%%%%%%%%%%%%%%%%%%%%%%%%%%%%%%%%%%%%%%%%%%%%%%%%%%%%%%%%

The astrophysically relevant problems of current carrying string loops in spherically symmetric spacetimes have been studied recently~\cite{Kol-Stu:2010:PHYSR4:, Stu-Kol:2012:PHYSR4:, Stu-Kol:2012:JCAP:}. In the present paper we investigate the relevant issues for Reissner-Nordstr\"{o}m background, giving the attention on the influence of black hole charge $Q$ and its electromagnetic interaction with string loop charge $\omega$ created by scalar field $\varphi$ living on the string loop.

Scalar field $\varphi$, living on the string loops and represented by the angular momentum parameter $J$, is essential for creating the centrifugal forces, and therefore for existence of stable string loop positions. In RN background is the charged string loop innermost stable equilibrium position (\ISSP) located between the photon circular orbit at $r_{\rm ph}$ and innermost stable charged particle orbit (ISCO). The condition $r_{\rm ph}<r_{\rm ISSP}<r_{\rm ISCO}$, already proven for rotating Kerr black hole background \cite{Kol-Stu:2013:PHYSR4:}, is supporting consideration of string loop model as a composition of charged particles and their electromagnetic fields~\cite{Cre-Stu:2013:PhRvE:}.

We have shown different types of string loop energy boundaries and different string loop trajectories in RN backgrounds. There are not any new types of the string loop motion for RN black hole background \cite{Jac-Sot:2009:PHYSR4:}, but in the field of RN naked singularities two closed toroidal regions for the string loop motion are possible ($Q=1.0677$, $\omega=1$, $J=10$) .

String loop harmonic oscillations around stable equilibria, defined by equation (\ref{fq2}), could be one of the perspective explanations of the HF QPOs observed in binary systems containing black holes or neutron starts. In the present paper, we applied the string loop resonant oscillations model to fit observed data from GRO 1655-40, XTE 1550-564, GRS 1915+105 microquasar sources. Our fittings are substantially compatible with the observed data from the GRO 1655-40 source and partially coincide with the GRS 1915+105 data. For the latter source the values of $\omega$ parameter are significant. Observed data from XTE 1550-564 can be explained only for $Q\sim0.9$ values. We can conclude that the twin HF QPOs could be efficiently explained by the string loop oscillatory model, if we consider interaction of an electrically charged current carrying axisymmetric string loop with the combined gravitational and electromagnetic fields of Kerr-Newman black hole where due to the combination of the black hole spin even small electric charge of the black hole can cause relevant modifications of the frequencies of the string loop oscillations.

String loop acceleration to the relativistic escaping velocities in the black hole neighbourhood, is one of the possible explanations of relativistic jets coming from AGN. We have studied the effect of black hole and string loop charges interaction on to the acceleration process. Due to the chaotic character of equations of motion, the positively charged string loop $\omega>0$ can be ejected from equatorial plane even in flat background.
The RN black hole charge $Q$ does not contribute to the string loop acceleration speeds due to electrostatic repulsion, it only modifies the effective potential $V_{\rm eff}$ and allows the string loop to came closer to the black hole, where the transmutation is more effective. This implies a surprising phenomena: the transmutation effect is more efficient, and the string loop is more significantly accelerated for the electric attraction of the string loop and the black hole, as the transmission process can occur in deeper regions of the gravitational potential well than in the case of electric attraction.
Note that contrary to the standard Blandford-Znajek mechanism of jet acceleration to high velocities \cite{Bla-Zna:1977:MNRAS:}, where fast rotating black hole must be assumed, in the string-loop acceleration model rotation of the black hole is nor required.

The RN solution is simple and elegant solution of combined Einstein and Maxwell equations and by studying charged string loop dynamics in this solution, we would like to just complete our previous string loop studies in order to map potential role of the Coulombic electric interaction. We explore theoretical properties of charged string loop motion in RN background and we show that unrealistically high values of RN charge are needed to explain real astrophysical data. In some dynamic situations, as those corresponding to unstable states of accretion disks wider ionization processes, the electric charge could be momentarily larger as indicated above for stationary situations.

\section*{Acknowledgments}

T.O. acknowledges the Silesian University in Opava Grant No. SGS/14/2016,  M.K. acknowledges the Czech Science Foundation Grant No. 16-03564Y and Z.S. acknowledges Albert Einstein Centre for Gravitation and Astrophysics supported by the Czech Science Foundation Grant No. 14-37086G.

\appendix

\section{Dimensional analysis and estimates of string loop parameters \label{appendix1}}
%%%%%%%%%%%%%%%%%%%%%%%%%%%%%%%%%%%%%%%%%%%%%%%%%%%%%%%%%%%%%%%%%%%%%%%%%%%

In the geometrized units the gravitational constant $G$ and the speed of light $c$ are taken to be dimensionless units. Their values and values of the Coulomb or electrostatic constant $k_{\rm e}$ and the Sun mass $M_{\sun}$ in the SI units are
\bea
 G &=& 6.67 \times 10^{-11} \,\, \rm{ {m^3}\cdot{kg^{-1} \cdot s^{-2}} }, \nn\\
 c &=& 3.00 \times 10^{8} \,\, \rm{ {m}\cdot{s^{-1}} }, \nn\\
 k_e &=& 8.99 \times 10^{8} \,\, \rm{ kg \cdot {m}^3 \cdot C^{-2} \cdot s^{-2} }, \nn\\
 M_{\sun} &=& 2.00 \times 10^{30} \,\, \rm{ kg }.
\eea

%The conversions of the fundamental quantities characterizing the string loop motion from the geometrized units to the Gaussian units are shown in the Table.~\ref{tabdim}. The table allows to perform transformation from the geometrized units to the CGS units, and vice versa, for any dynamical quantity describing the string loop dynamics.

For central \Schw{} black hole with $M=10 M_\sun$, black hole length scales can be calculated in SI units
\bea
r_{\rm hor}=2 M G / c^2 = 3 \times 10^{4} \,\, \rm{ m }, \nn \\
r_{\rm loop}=2 \pi \cdot 6 M G / c^2 = 5.6 \times 10^{5} \,\, \rm{ m }, \label{ourBH}
\eea
where $r_{\rm hor}$ \Schw{} horizon radius (black hole size) and $r_{\rm loop}$ is length of loop located at ISCO (inner edge of Keplerian accretion disc). For \Schw{} or RN black hole, the radial coordinate $r$ is circumferential.

\subsection{Astrophysical relevance of black hole charge}

One can compare the characteristic length scale given by the charge of the RN black hole $Q_{\rm G}$ with its gravitational radius. This gives the charge, whose gravitational effect is comparable with the spacetime curvature of a black hole. For the black hole of mass $M$ this condition implies that the gravitational effect of the charge $Q$ on the background geometry can be neglected if
\beq
Q << Q_{\rm G} = 2 \sqrt{ \frac{G}{k_{\rm e}} } M \approx 10^{20} \frac{M}{M_\sun} \, \rm{C}. \label{QG}
\eeq
If $Q<<Q_{\rm G}$, the electric field cannot modify the background geometry of the black hole, but still there can be electrostatic interaction between black hole and particle/string charges.

Reissner-Nordstr\"{o}m black hole charge $Q$ is assumed to be small or even negligible for realistic black holes. Since gravitation interaction is quite weak compared to the electromagnetic interaction, with ration $e/\sqrt{G}m_{\rm p} \sim 10^{18}$, any RN charged black hole will easily separate electrons and protons from surrounding plasma and neutralizing the RN black hole with charge $Q>10^{-18} Q_{\rm G}$ quickly \cite{Ear-Pres:1975:ARAA:}. The amount of material, necessary for neutralization of maximally charged RN black hole $Q=Q_{\rm G}$ is small
\beq
 M_{\rm acretion} \sim 10^{-18} M \sim 10^{12} \frac{M}{M_\sun} \, {\rm kg}.
\eeq

\subsection{String loop parameters}

The string loop model enables to apply and compare the derived solutions for different physical mechanisms to obtain the estimates of the parameters characterizing the string loop dynamics. In order to make the estimate of the tension $\mu$ strength, one can use, e.g., the similarity between the role of the parameter $\mu$ and the Lorentz force acting on a charged particle in the action governing the string loop dynamics. On the other hand, we can find estimates of the fundamental string loop parameter values related to the so called cosmic strings, giving the upper limit of the application of the string loop model. Realistic estimations give in the SI units the string loop tension $\mu$ in order \cite{Tur-etal:2014:PHYSR4:}
\bea
\mu_{({\rm L})}\leq\mu<\mu_{({\rm CS})}, \qquad \mu_{({\rm L})} = 10^{-14} {\rm kg/m},\\
\mu_{({\rm CS})} < 10^{20} {\rm kg/m}.
\eea

To prove that string loop is test object only, we give some examples of string loop mass and charge. Total string loop mass $m_{\rm loop}$ will be related to the total string loop energy $E_{\rm loop}$ by mass-energy equivalence formula
\beq
 E_{\rm loop} = m_{\rm loop} c^2.
\eeq
For the Nambu-Goto string loop with radius $r$ the total string energy is just string length $2\pi r$ times tension $\mu$, giving for string loop mass formula
\beq
 m_{\rm loop} = 2 \pi r \mu / c^2.
\eeq
For our choice (\ref{ourBH}) of black hole $M=10 M_{\rm sun}$, we have extremely light string loop in Lorentz case $ m_{\rm loop(L)} \sim 10^{-10} \rm{kg}$, while "Earth mass" loop $m_{\rm loop(CS)} \sim 10^{24} \rm{kg}$ in cosmic string case.

We can give ratio between the total mass of the charged string loop $m_{\rm loop}$ and the mass of RN black hole $M$, and ratio between string loop charge $q_{\rm loop}$ and charge of RN black hole $Q$ in SI units by formulas
\beq \label{formulaM}
\frac{m_{\rm loop}}{M} = 2\pi \, \frac{\mu G}{c^4} \, E, \qquad \frac{q_{\rm loop}^2}{Q^2} = \frac{4 \pi^2}{a^2} \, \frac{\mu G}{c^4} \, \Omega^2,
\eeq
where $E$ and $\Omega$ are previously used dimensionless string loop energy and charge density and $a$ is ratio between charge and mass of the RN black hole $a = Q/M \in (0,1)$. For charged string loop $\omega = 1$ at stable position $r=6$ around RN black hole with $a = 0.5$, the dimensionless parameters are $ E \doteq 13, \Omega \doteq 10$. Since the term $ G \mu / c^{4} $ is very small, $10^{-31}$ for Lorentz and $10^{-7}$ for cosmic strings, the string loop total mass $m_{\rm loop}$ and total charge $q_{\rm loop}$ are negligible in comparison the the RN black hole mass $M$ and charge $Q$. Only if $a \rightarrow 0$ then RN black hole charge $Q$ will become comparable to the charge of string loop $q_{\rm loop}$.

%%%%%%%%%%%%%%%%%%%%%%%%%%%%%%%%%%%%%%%%%%%%%%%%%%%%%%%%%%%%%%%%%%%%%%%%%%%%%%%%%%%%%%%%%%%%%%
%\section*{References}

%\section*{References}

\def\prc{Phys. Rev. C}
\def\pre{Phys. Rev. E}
\def\prd{Phys. Rev. D}
\def\jcap{Journal of Cosmology and Astroparticle Physics}
\def\apss{Astrophysics and Space Science}
\def\mnras{Monthly Notices of the Royal Astronomical Society}
\def\apj{The Astrophysical Journal}
\def\aap{Astronomy and Astrophysics}
\def\actaa{Acta Astronomica}
\def\pasj{Publications of the Astronomical Society of Japan}
\def\apjl{Astrophysical Journal Letters}
\def\pasa{Publications Astronomical Society of Australia}
\def\nat{Nature}
\def\physrep{Physics Reports}
\def\araa{Annual Review of Astronomy and Astrophysics}
\def\apjs{The Astrophysical Journal Supplement}
\def\aapr{The Astronomy and Astrophysics Review}
\def\procspie{Proceedings of the SPIE}

%\bibliographystyle{plain}
%\input{C:/DOC/WORK/TXT/reference/refdef}
%\bibliography{C:/DOC/WORK/reference/reference}

\end{document}